\documentclass[traditabstract]{aa}
\usepackage{txfonts}
\usepackage{graphicx}
\usepackage{natbib}
\usepackage{aalongtable}

\bibpunct{(}{)}{;}{a}{}{,}

\begin{document}

\title{Period and period change measurements for 143 SuperWASP\\eclipsing binary candidates near the short-period limit\\and discovery of a doubly eclipsing quadruple system}
\author{M.~E.~Lohr\inst{\ref{inst1}}\and A.~J.~Norton\inst{\ref{inst1}}\and U.~C.~Kolb\inst{\ref{inst1}}\and P.~F.~L.~Maxted\inst{\ref{inst2}}\and I.~Todd\inst{\ref{inst3}}\and R.~G.~West\inst{\ref{inst4}}}
\institute{Department of Physical Sciences, The Open University,  Walton Hall, Milton Keynes MK7\,6AA, UK\\ \email{Marcus.Lohr@open.ac.uk}\label{inst1}\and Astrophysics Group, Keele University, Staffordshire ST5\,5BG, UK\label{inst2}\and Astrophysics Research Centre, School of Mathematics \& Physics, Queen's University, University Road, Belfast BT7\,1NN, UK\label{inst3}\and Department of Physics and Astronomy, University of Leicester, Leicester LE1\,7RH, UK\label{inst4}}
\date{Received 15 October 2012 / Accepted 19 November 2012}

\abstract {Building on previous work, a new search of the SuperWASP
  archive was carried out to identify eclipsing binary systems near
  the short-period limit.  143 candidate objects were detected with
  orbital periods between 16\,000 and 20\,000~s, of which 97 are new
  discoveries.  Period changes significant at $1\sigma$ or more were
  detected in 74 of these objects, and in 38 the changes were
  significant at $3\sigma$ or more.  The significant period changes
  observed followed an approximately normal distribution with a
  half-width at half-maximum of $\sim$0.1~s~yr\textsuperscript{-1}.
  There was no apparent relationship between period length and
  magnitude or direction of period change.  Amongst several
  interesting individual objects studied,
  \object{1SWASP~J093010.78+533859.5} is presented as a new doubly
  eclipsing quadruple system, consisting of a contact binary with a
  19\,674.575~s period and an Algol-type binary with a 112\,799.109~s
  period, separated by 66.1~AU, being the sixth known system of this
  type.}

\titlerunning{143 SuperWASP eclipsing binaries including doubly eclipsing system}
\authorrunning{M.~E.~Lohr et al.}

\maketitle

\section{Introduction}

The SuperWASP (Wide Angle Search for Planets) project \citep{pollacco}
has conducted wide-field time-domain photometric surveys since 2003 in
the northern hemisphere, and since 2005 in the southern hemisphere.
Employing 30~s exposures and 11~cm telescope apertures, its emphasis
has been on obtaining high-cadence (6--40~minute) observations of
bright ($V\sim8$--15~mag) stars over almost the whole sky, with a
primary goal of identifying transiting exoplanet candidates.  Although
it has been highly successful in this (69 exoplanets announced by
early October 2012, out of
287\footnote{http://exoplanet.eu/catalog.php}), it is also well-suited
to the detection and study of variable stars of many types, and in
particular eclipsing binaries.

Here we have used the SuperWASP archive of $\sim$30 million objects to
search for and analyse main sequence eclipsing binaries with very
short orbital periods ($<$20\,000~s or $\sim$0.2315~d).  This should
yield an interesting sample around the observed short-period limit for
such binary systems of $\sim$0.2~d \citep{ruc92, ruc07}, potentially
illuminating the causes of this cut-off point in the period
distribution.  (We should note, however, that systems in this period
range must be expected to be relatively low in mass and hence
intrinsically faint, such that SuperWASP will not detect them with the
same efficiency as longer-period eclipsing binaries.)  In earlier
work, \citet{norton} presented 53 candidate eclipsing systems in this
period range, using SuperWASP archived data; here, with a more
thorough search, we have sought to detect further such objects which
might have been missed.  Also, \citet{lohr} presented the results of a
search for period changes in these 53 objects, finding three which
exhibited statistically significant period decrease; here again, we
have used an improved period change detection method to search for
period changes in the eclipsing systems found with periods below
20\,000~s.  This paper, then, is primarily intended to update
\citet{norton} and \citet{lohr}.  However, in the process of analysing
our findings, we believe we have discovered a new doubly eclipsing
quadruple system, which is reported in Subsect.~\ref{sub:J093010}

\section{Method}
\label{method}

An initial list of 36\,758 SuperWASP identifiers was obtained from the
catalogue, with associated possible periods in the range
8000--10\,000~s.  This range would correspond to potential binary
orbital periods below 20\,000~s, since there are two eclipses per
cycle.  (The catalogue periods are the result of a uniform period
search applied to the majority of data as part of the initial
processing pipeline; the code used is described in \citet{norton07}
and is run separately for data from different seasons and cameras.  As
a consequence, several different periods can be listed in the
catalogue for a single object.)  A frequency plot of the periods
revealed substantial excesses of objects in the ranges 8610--8625~s
and 9565--9586~s i.e. in the neighbourhood of 1/10 and 1/9 of a
sidereal day respectively.  Since the vast majority of these harmonic
periodicities are expected to be spurious, objects in these ranges
were excluded from further consideration.  It is likely that only
about three genuine eclipsing binaries with periods below 20\,000~s
will have been missed as a consequence.  Repeated identifiers
(occurring when a single identifier had multiple possible
periodicities listed in the catalogue, in the ranges of interest) were
also removed at this stage.

This left 5743 distinct identifiers, of which around 5190 probably
represented distinct astrophysical sources: since SuperWASP uses the
USNO-B1 input catalogue to label objects, it is possible for a single
bright source in a field of view containing many faint sources to be
catalogued under multiple identifiers.  Such `duplicates' can usually
be easily recognised by their near-identical periods and coordinates;
however, they were not excluded at this stage so that the brightest or
clearest lightcurve for each source could be preferentially selected
for more detailed analysis later.  Lightcurves with fluxes corrected
by the Sys-Rem algorithm \citep{tamuz,mazeh} were obtained from the
archive for these objects.

A custom-written IDL program was then used to check and refine the
catalogue periods in a two-step process which improved on that used in
\citet{lohr}.  First, up to 50 short sections (depending on file
length) of each lightcurve were used for trial fitting with a
sinusoidal function, using the Levenberg-Marquardt algorithm
\citep{lev,marq}.  If a frequency plot of the resulting periods
yielded a single dominant approximate period, this was used as the
period estimate for the second step; otherwise the object was
classified as probably non-periodic or possessing a period outside the
8000--10\,000~s range, and not considered further.  4434 objects of
interest remained after this step.

The approximate periods found for these objects were then refined to
the nearest 0.001~s by a form of phase dispersion minimization
i.e. recursively folding the lightcurve on trial periods separated by
small intervals, and selecting the period which minimized the standard
deviation of flux values in each of 100 phase bins.  This step was
repeated with the initial trial period being doubled, and objects were
given a preliminary classification as possible eclipsing binaries if
the minimum phase dispersion was lower with the doubled period than
with the single period; the doubled period was then retained as the
binary's orbital period.  If the single period yielded a lower minimum
phase dispersion, the object was tentatively classified as a periodic
variable of a different type (probably a pulsating or rotating
variable, in this period range, or a non-eclipsing contact binary),
and the single period was retained as the star's pulsational or
rotational period.

Since this method will not always separate pulsators and eclipsing
binaries reliably (e.g. in cases where eclipsing systems show primary
and secondary eclipses of equal depth), a final visual check was made
of the folded lightcurves of the more distinctive objects with periods
below 20\,000~s.  This included all those where the amplitude of the
mean lightcurve exceeded the amplitude of data scatter about the mean
lightcurve, facilitating a clear identification of variable type.  It
also included some where the amplitude of variation was between 50\%
and 100\% of the scatter, but where the objects might be expected to
have distinctive lightcurves on other grounds e.g. very bright
objects; objects with high numbers of observations; objects with very
well-defined periods found in step one of the program.

Approximately 1000 identifiers were checked, and 201 were selected as
probable eclipsing binaries of W~UMa type, corresponding to 143
distinct astrophysical objects.  Fainter duplicates were rejected at
this stage.  The selected objects had nearly all been automatically
classified as eclipsing binaries, and many showed a clear difference
in depths of primary and secondary eclipses; some had different
heights of maxima, presumably due to the O'Connell effect
\citep{oconnell}; others were chosen for their relatively broad,
symmetric maxima and narrow eclipses.  Purely sinusoidal lightcurves
were excluded; although some of them were probably generated by
genuine eclipsing binaries, our photometric data was insufficient to
distinguish them from other periodic variable types with reasonable
probability.  Radial velocities would need to be determined for these
ambiguous objects to establish their variability type.

Evidence of period change was then searched for in these 143 objects
by means of automated construction of O$-$C (observed minus
calculated) diagrams, using an improvement of the method described in
\citet{lohr}.  The main difference was in the method for determining
accurate times of observed (primary) minima.  Rather than trying to
fit local regions of the lightcurve with quadratic, Gaussian,
sinusoidal or other analytic functions, the binned mean lightcurve for
each object, found during phase dispersion minimization, was used as a
fitting `function'.  This shape, being derived from the combined
observations of hundreds of cycles, might be expected to provide an
excellent fit for each individual observed cycle, since it represents
the true underlying shape of the object's lightcurve.  A similar
approach was used by \citet{pribulla08} for finding eclipse minima in
a close quadruple system.

This approach has the advantages that it does not require the eclipses
to be remotely symmetrical, or the same phase range to be used for
each fit, as is preferred with the method of \citet{kwee}.  The method
might be expected to break down, however, in objects whose lightcurve
varies in shape over time, such as RR~Lyrae pulsators exhibiting the
Blazhko effect \citep{blazhko}.  The number of phase bins to be used
for the fitting curve needs to be chosen carefully: if too many bins
are used, the fitting curve will appear spiky, with features which are
not part of the underlying lightcurve; if too few bins are used, the
fitting curve will blur distinctive features of the underlying
lightcurve which may be necessary for optimal fits e.g. a difference
in depth or shape between primary and secondary eclipses.  For this
study, the program automatically picked a bin number based on the
number of data points in the object file and the brightness of the
object, but this was optimized by hand in some cases.  An optimal
night for fitting the zero epoch (essential for determining reliable
calculated times of eclipses for the O$-$C diagram) was also selected
manually from a range near the middle night of each object file.

After construction of each O$-$C diagram, a small number of
locally-outlying values were stripped out automatically; manual
checking of the corresponding nights of observation indicated that
these were typically caused by irregular features of the data
(instrumental or astrophysical) rather than failures of the fitting
method.  Linear and quadratic fits of the O$-$C values were then
attempted; where a quadratic function gave a superior fit, and the
rate of change was significantly different from zero ($\ge 1\sigma$),
the object was counted as exhibiting secular period change.

\section{Results}

During the checking process described above, periodic variables of
several different types were observed.  A large number of pulsators
(probably $\delta$~Scuti and RR~Lyrae variables in this short-period
range) were evident, characterized by narrow minima and broad maxima,
and/or asymmetric lightcurve shapes.  As well as eclipsing binaries in
the period range of interest, many eclipsing binaries were seen with
periods between 32\,000 and 40\,000~s, whose quarter-periods had
evidently been picked up by the period detection code used in the
SuperWASP catalogue.  Also, the subdwarf B (sdB) eclipsing binaries
\object{NY~Vir}, \object{HS~2231+2441} and \object{NSVS~14256825},
which have orbital periods in the range 8000--10\,000~s, were detected
and phase-folded appropriately in spite of their strongly
non-sinusoidal lightcurves.  These detections suggest that the period
determination method used here, despite using sinusoidal fitting
during the initial stage, is effective with periodic variables
exhibiting a wide variety of lightcurve shapes.

\subsection{Periods}

\begin{figure}
\resizebox{\hsize}{!}{\includegraphics{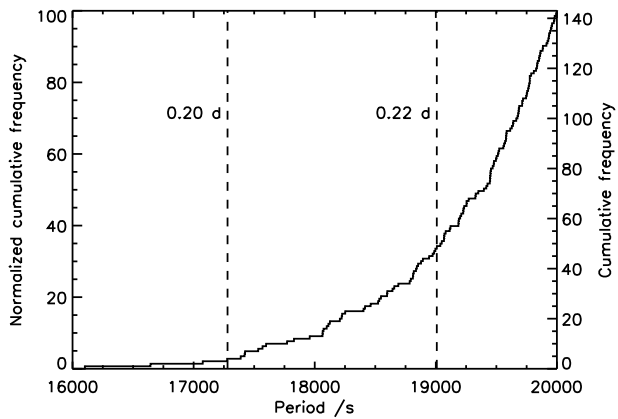}}
\caption{Cumulative period distribution of 143 candidate eclipsing binaries.
Typical values for the short-period limit are shown.}
\label{perioddist}
\end{figure}

Period and magnitude statistics for the 143 candidate (main sequence)
eclipsing binaries found with periods under 20\,000~s are presented in
Table~\ref{143data} (online only).  They include 44 of the 53
short-period candidate eclipsing binaries presented in \citet{norton}.
(The other nine, as discussed in \citet{lohr}, have periods slightly
longer than 20\,000~s.)  Three of the 99 additional objects presented
here are known periodic variables: \object{CC~Com}, listed as a
W~UMa-type eclipsing binary in the GCVS, with the same period as found
here; \object{LL~Eri}, listed as a rotating ellipsoidal binary in the
GCVS, again with the period found here; and
\object{ROTSE1~J164349.58+325637.8}, listed in \citet{akerlof} as a
$\delta$~Scuti variable with period half that found here.  Since the
SuperWASP lightcurve for the latter object is particularly
well-observed, and its folded lightcurve exhibits clear (though small)
differences in the depths of primary and secondary minima, we suggest
that the object is more likely to be an eclipsing binary.  Therefore,
97 of the objects presented here are new candidate eclipsing binaries
near the short-period limit.  Figure~\ref{perioddist} gives their
cumulative period distribution, and Fig.~\ref{appfig1} (online only)
shows their individual folded lightcurves.

\subsection{Period changes}
\label{pdots}

\begin{figure}
\resizebox{\hsize}{!}{\includegraphics{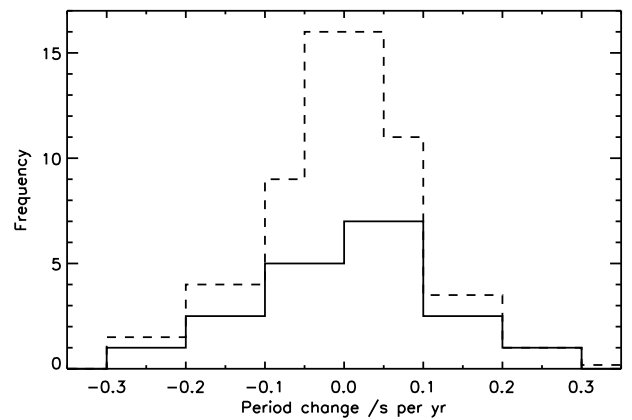}}
\caption{Period change distribution of 74 candidate eclipsing binaries
  showing significant change.  Period changes found with at least
  $3\sigma$ confidence are indicated with solid lines (38 objects);
  dashed lines indicate changes found at a $1\sigma$ level or more.}
\label{periodchangedist}
\end{figure}

12 objects were excluded from the period change search, since only a
single year of data was available for them.  Of the remaining 131, 74
showed evidence of significant secular change i.e. the uncertainty
ranges on their $\dot{P}$ values did not include zero.  (Note that
highly significant period change may be small in magnitude: J121206
shows a change of just 0.0265~s~yr\textsuperscript{-1}, but its
remarkably bright and well-defined lightcurve makes this value
significant at $23\sigma$.)  The remainder have not necessarily been
demonstrated to have unchanging periods: although the O$-$C diagrams
of some were indeed better fitted by a linear function than a
quadratic one, others exhibited apparent secular change at a
non-significant level, and may be regarded as having indeterminate
status.  Period changes and significance levels are listed for these
74 objects in Table~\ref{143data} (online only), and the distribution
of significant changes is plotted in Fig.~\ref{periodchangedist}.
Figure~\ref{appfig1} (online only) shows the O$-$C diagrams of all
objects.

Since a few of the changes significant at 1 or $2\sigma$ may be
expected to be spurious, resulting by chance alone in a data set of
this size, the distribution of changes significant at $\ge3\sigma$ is
also shown in Fig.~\ref{periodchangedist} (38 objects); we would not
expect to find even one instance of period change this significant out
of 131 objects by chance.  Two of the objects found to exhibit highly
significant period change in \citet{lohr} are among this set
(J133105 and J234401); the third (1SWASP~J174310.98+432709.6)
has a period slightly greater than 20\,000~s and so was not included
in the present study.  The reproduction of our earlier findings
supports the validity of the modified period change detection
approach, and the increased number of highly-significant period
changes detected implies an improvement in sensitivity.

\section{Discussion}

\subsection{Periods}

It may be seen that the period distribution found here
(Fig.~\ref{perioddist}) straddles two frequently-quoted values for the
period cut-off (0.20 and 0.22~d), and follows a smooth tail-off
towards shorter periods.  It tallies well with similar distributions
found from other wide-field time-domain surveys, covering wider period
ranges, such as those in \citet{szymanski}, Fig.~9, and
\citet{paczynski}, Fig.~6, which show contact binary periods tailing
off towards lower periods from a maximum around 0.38~d, with none
observed below 0.20~d.

Only one object (J201816) is included here with a period shorter
than that of \object{BX~Tri} (=GSC~02314-0530, here J022050),
generally regarded as the main-sequence eclipsing binary with the
shortest-known period \citep{dimitrov}.  However, since this object is
very poorly-observed by SuperWASP, with only a couple of thousand data
points mostly from a single year, we are reluctant to make too strong
a claim for it.  Although it apparently exhibits substantial
differences in the heights of its maxima, as well as small differences
in primary and secondary minima depths, it is possible that these are
artefacts of the limited observations, and that the object is really a
pulsating variable.

Thus our search supports the existence of a short-period limit at
around 0.20~d, and perhaps favours particular types of explanation for
it.  \citet{stepien} argued that the current age of the Universe
indirectly explained the limit, on the grounds that lower-mass
detached binaries lose angular momentum more slowly than high-mass
systems, and so take longer to evolve into (stable) contact
configurations; 0.20~d would then simply be the current minimum period
that a system would have had time to reach.  Such a model might imply
a much sharper, more cliff-like cut-off point for the binary period
distribution than seen here (and indeed, a cut-off period which would
become shorter as the Universe aged).  Also, \citet{jiang} have
indicated a number of known short-period binaries with measured masses
lower than \citeauthor{stepien}'s formula should allow, and
\citet{nefs} have claimed the discovery of four M-dwarf eclipsing
binaries with periods below 0.18~d (one as short as 0.11~d), which
would certainly conflict with \citeauthor{stepien}'s model.

However, other models suggest that objects are leaving the
short-period end of the distribution through rapid merger, in addition
to entering it from above through evolution from detached into contact
configurations.  \citet{jiang} argued that binary systems with
particular combinations of low primary mass and low mass ratio would
evolve into unstable states and merge rapidly; 0.20~d would then be
the shortest period corresponding to a possible stable configuration.
\citet{stepien12} also proposes a new series of binary models
including evolution towards coalescence within the contact stage; the
lowest period obtained for any of these models, at the time of
coalescence, is 0.201~d.  Such explanations might fit better with the
observed distribution of binary periods, with some objects reaching
unstable states and heading rapidly towards merger at periods somewhat
above 0.20~d, and others able to remain stable even at the cut-off
point.

\subsection{Period changes}

\begin{figure}
\resizebox{\hsize}{!}{\includegraphics{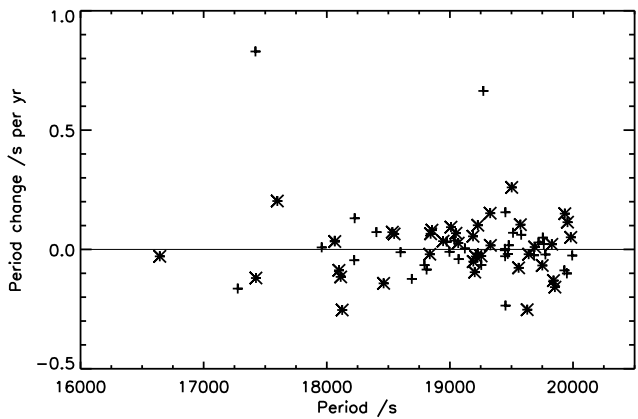}}
\caption{Plot of significant period changes against periods, for 74
objects significant at $\ge1\sigma$ (small crosses) and 38 objects
significant at $\ge3\sigma$ (larger diagonal crosses).}
\label{pvpdot}
\end{figure}

The distribution of significant period changes observed here
(Fig.~\ref{periodchangedist}) also tallies broadly with one found by
\citet{kubiak}, Fig.~5, for 134 OGLE contact binaries with periods
below 1~d.  It is symmetrical around zero (a Kolmogorov-Smirnov test
supports symmetry at $P=0.91$ for the 74 objects with period change
significant at $\ge1\sigma$ and at $P=0.69$ for the 38 objects
significant at $\ge3\sigma$), and approximately normal, with a
half-width at half-maximum around 0.1~s~yr\textsuperscript{-1}, where
\citeauthor{kubiak} found a rather lower value around
0.03~s~yr\textsuperscript{-1}.  The extremely short-period objects
considered here, then, appear equally likely to show increases and
decreases in period.  Figure~\ref{pvpdot} plots significant period
changes against periods.  No particular relation between either
direction or magnitude of period change and period length seems
apparent.

It might be argued that the period change distribution found here has
an effective hole at zero, and a deficiency near zero, since we do not
include objects where no significant change was detected.  As
indicated in Subsect.~\ref{pdots}, these objects are not of a single
type, and cannot be confidently claimed to have unchanging periods.
To include all or some of them in our distribution would distort it
and make it hard to compare with other results obtained using
different methods.  \citeauthor{kubiak}'s distribution of objects
(with $\mathcal{P}$-statistic $>63.3$, described as ``statistically
confirmed'' period changes) also has a gap near zero, for similar
reasons to ours: it is more difficult to detect and quantify small
period changes since their uncertainties must also be small.

Another potential confounding factor would be the presence of
wandering spots on the surface of a star.  These could in theory
change the lightcurve shape in such a way as to move the detected
times of minima and so create a spurious curvature in the O-C diagram,
leading us to conclude erroneously that the object's orbital period is
changing.  In practice, visual checks of individual nights suggest the
lightcurves do not vary substantially in shape for the objects showing
the most significant period change, though it is possible that spot
movement is contributing to data scatter in some of the fainter
and less well-observed lightcurves.

Some of the O$-$C diagrams suggest a periodic, sinusoidal variation in
period, in addition to or instead of a secular trend (e.g. J142312,
J172717, J210423).  Such cyclical variations are often interpreted
as indicating the presence of a third body in the binary system
(e.g. \citet{lee}), through the light-time effect.  However, the
Applegate mechanism \citep{applegate}, involving the magnetic activity
cycle of a star in a close binary system, may often provide a more
plausible explanation \citep{hilditch,christopoulou}; this mechanism
explains orbital period changes as gravitational quadrupole responses
to a cyclic redistribution of angular momentum within the layers of an
active, convective star, associated with varying levels of
differential rotation at different times in the star's magnetic cycle.
In such a case, longer-term luminosity variations would be expected to
show the same period as the orbital period modulation, since they
arise from a common cause, and any other variability associated with
magnetic activity e.g. coronal X-ray luminosity, should also show this
period.

\subsection{J234401}

\begin{figure}
\resizebox{\hsize}{!}{\includegraphics{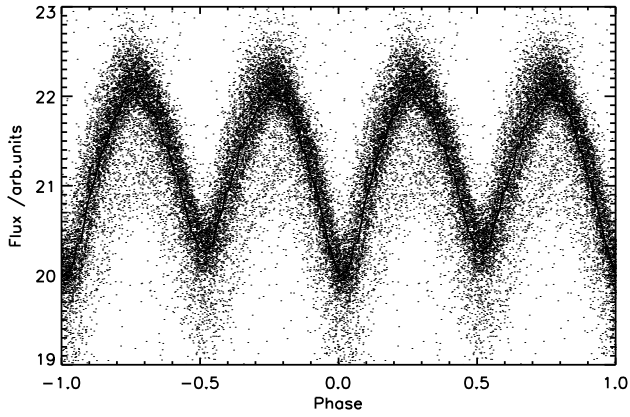}}
\caption{Lightcurve of object J234401 folded at period 18\,461.639~s.}
\label{J234401lc}
\end{figure}

\begin{figure}
\resizebox{\hsize}{!}{\includegraphics{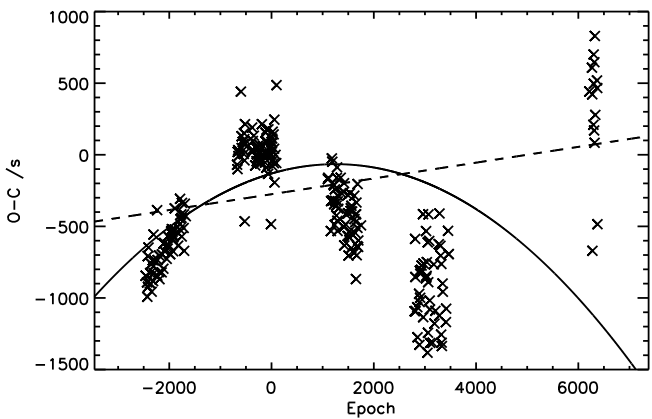}}
\caption{O$-$C diagram for object J234401 (uncertainties not plotted
for clarity).  Dashed line shows best linear fit ($\chi^2=22.53$);
solid line shows best quadratic fit ($\chi^2=17.83$), corresponding to
a secular period change of $-0.1422\pm
0.0041$~s~yr\textsuperscript{-1}.}
\label{J234401oc}
\end{figure}

\begin{figure}
\resizebox{\hsize}{!}{\includegraphics{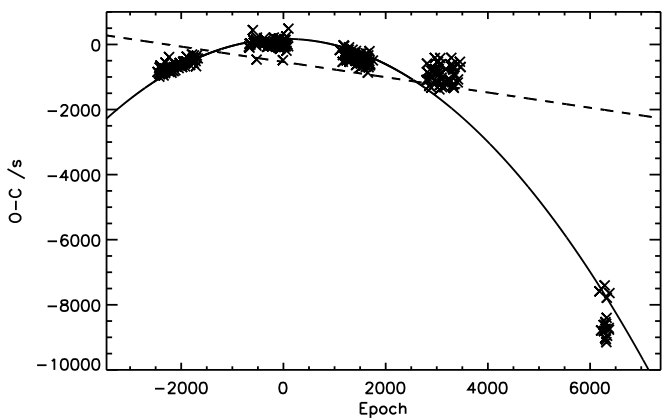}}
\caption{O$-$C diagram for object J234401 with final year times of
minima adjusted by half a cycle, to test the hypothesis that secondary
minima are closer to the calculated times than primary minima, owing
to rapidity of period decrease.  Dashed line shows best linear fit
($\chi^2=124.67$); solid line shows best quadratic fit
($\chi^2=10.86$), corresponding to a secular period change of
$-0.6902\pm 0.0040$~s~yr\textsuperscript{-1}.}
\label{J234401oc2}
\end{figure}

J234401 (Fig.~\ref{J234401lc}) may provide an example of such cyclical
variation.  In \citet{lohr}, we advanced this system as illustrating
very rapid period decrease ($-0.313\pm0.019$~s~yr\textsuperscript{-1},
significant at $16\sigma$) on the basis of the first four years of
SuperWASP observations.  However, with the addition of a further
partial year, following a year-long gap in observations of the object,
the O$-$C diagram now presents a rather different picture
(Fig.~\ref{J234401oc}).  Highly significant period change is still
indicated (at $35\sigma$ now), but its magnitude is weakened by the
addition of the further data, and the quadratic fit is evidently a
poor match.  One possible explanation is that the object is in fact
varying in period sinusoidally, with a period of around 4--4.5 years.
Such cyclical period variation could itself be superimposed on a
secular change, the direction of which would require a longer
time-base of observations to establish.

If the period variation is sinusoidal, and due to the Applegate
mechanism, the object would have a (semi-amplitude) $\Delta P/P$ of
$\sim3.3\times$10\textsuperscript{-5}, according to
\citeauthor{applegate}'s equation (38).  This is at the upper end of
period modulation amplitudes considered by \citeauthor{applegate} as
explicable by his mechanism, but of the same order of magnitude.
There is also a possible suggestion of average flux variation from
year to year, which could support the Applegate mechanism as the
explanation here.  However, if the sinusoidal variation were due to a
third body, this object would have to be of comparable mass to the
binary system itself (using equations 10 and 11 in \citet{pribulla},
and estimates for binary parameters described in \citet{lohr}).  With
a separation of $\sim$3~AU between third body and binary, we would not
expect to be able to resolve its light separately from the contact
system, but whether such a triple system would be stable in the long
term is unclear, and would require modelling to assess.

Alternatively, J234401's period could have continued to decrease
after the first four years, with the O$-$C values for the final year
being so large and negative that they have been `wrapped round' by
half a cycle i.e. the secondary eclipses now occur near to where the
primary eclipses would be calculated to occur on the basis of an
unchanging period.  Since the primary and secondary eclipses are of
similar depth and shape, this would not create a problem for the
fitting algorithm.  Figure~\ref{J234401oc2} indicates the O$-$C
diagram that would result if this had in fact occurred.  A quadratic
function now provides a far better fit, though an acceleration of
period decrease would be indicated
($-0.6902\pm0.0040$~s~yr\textsuperscript{-1}) and the data for year 4
now stands out as discrepant.  Period determinations for the
individual years of data also do not support a continuing rapid
decrease, though the uncertainties in period are substantial for the
final year, which contains relatively few nights of observations.

What is clear, however, is that J234401 is undergoing highly
significant and dramatic period changes of some sort.  We hope in the
near future to use newly-obtained spectroscopic data to learn more
about this interesting object.

\subsection{J102328}

\begin{figure}
\resizebox{\hsize}{!}{\includegraphics{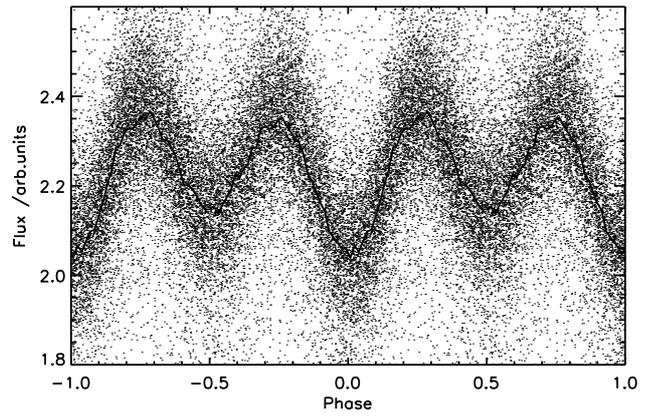}}
\caption{Lightcurve of object J102328 folded at period 18\,125.146~s.}
\label{J102328lc}
\end{figure}

\begin{figure}
\resizebox{\hsize}{!}{\includegraphics{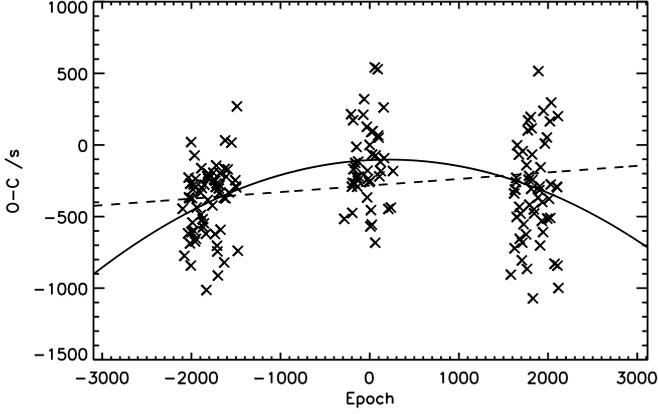}}
\caption{O$-$C diagram for object J102328 (uncertainties not plotted
for clarity).  Dashed line shows best linear fit ($\chi^2=1.74$);
solid line shows best quadratic fit ($\chi^2=1.46$), corresponding to
a secular period change of $-0.254\pm
0.037$~s~yr\textsuperscript{-1}.}
\label{J102328oc}
\end{figure}

Another object of note is J102328 (Figs.~\ref{J102328lc} and
\ref{J102328oc}), which shows period decrease nearly as rapid as that
apparently seen in J234401 on the basis of its first four years of
data: $-0.254\pm0.037$~s~yr\textsuperscript{-1}, significant at
$6\sigma$.  If this decrease continued it would imply a merger
timescale ($P/\dot{P}$) of at most 71\,500 years; however, only three
years of SuperWASP observations are available for this object, so
caution is warranted: future years of data might support a sinusoidal
period variation instead.  We may also note that this magnitude of
period decrease falls within the symmetrical distribution found for
the 38 highly significant objects taken as a group.

\subsection{J093010}
\label{sub:J093010}

\begin{figure}
\resizebox{\hsize}{!}{\includegraphics[trim = 7mm 0mm 0mm 0mm,
clip,angle=270]{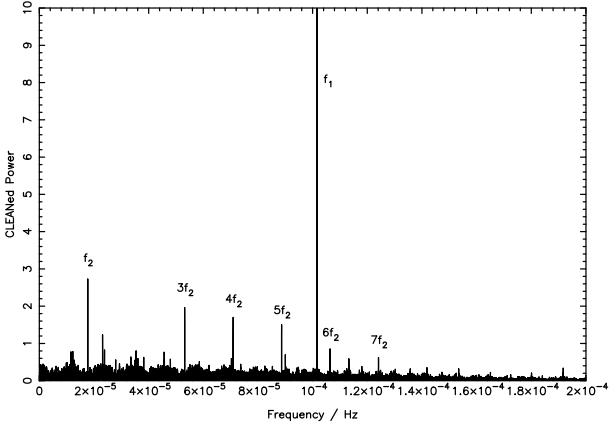}}
\caption{Power spectrum for J093010, showing strongest signal at $f_1$
(associated with the period of the contact binary) and a weaker signal
at $f_2$ and its harmonics (associated with the period of the
Algol-type binary).}
\label{powerspec}
\end{figure}

\begin{figure}
\resizebox{\hsize}{!}{\includegraphics{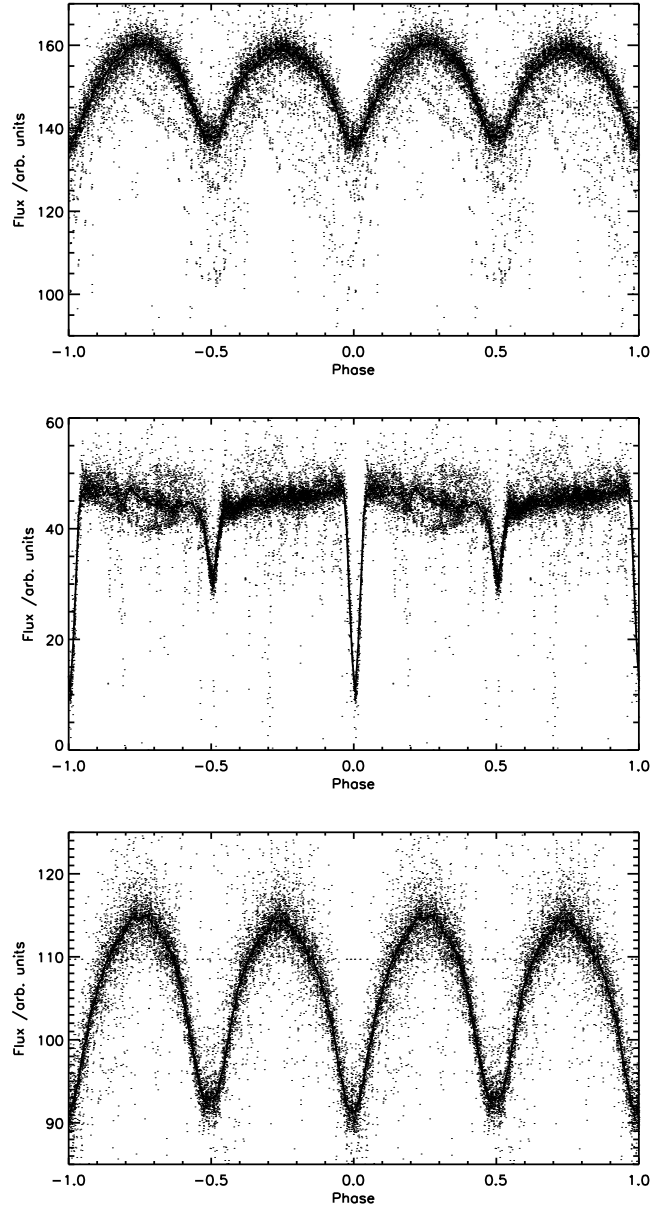}}
\caption{Top: lightcurve of object J093010 (combined with duplicate)
folded at dominant contact binary period 19\,674.574~s, with median
binned lightcurve overplotted.  Middle: lightcurve of Algol-type
eclipsing binary folded at period 112\,799.109~s, after subtraction of
median binned lightcurve.  Bottom: lightcurve of W~UMa-type eclipsing
binary after subtraction of Algol median binned lightcurve.}
\label{J093010}
\end{figure}

\begin{figure}
\resizebox{\hsize}{!}{\includegraphics{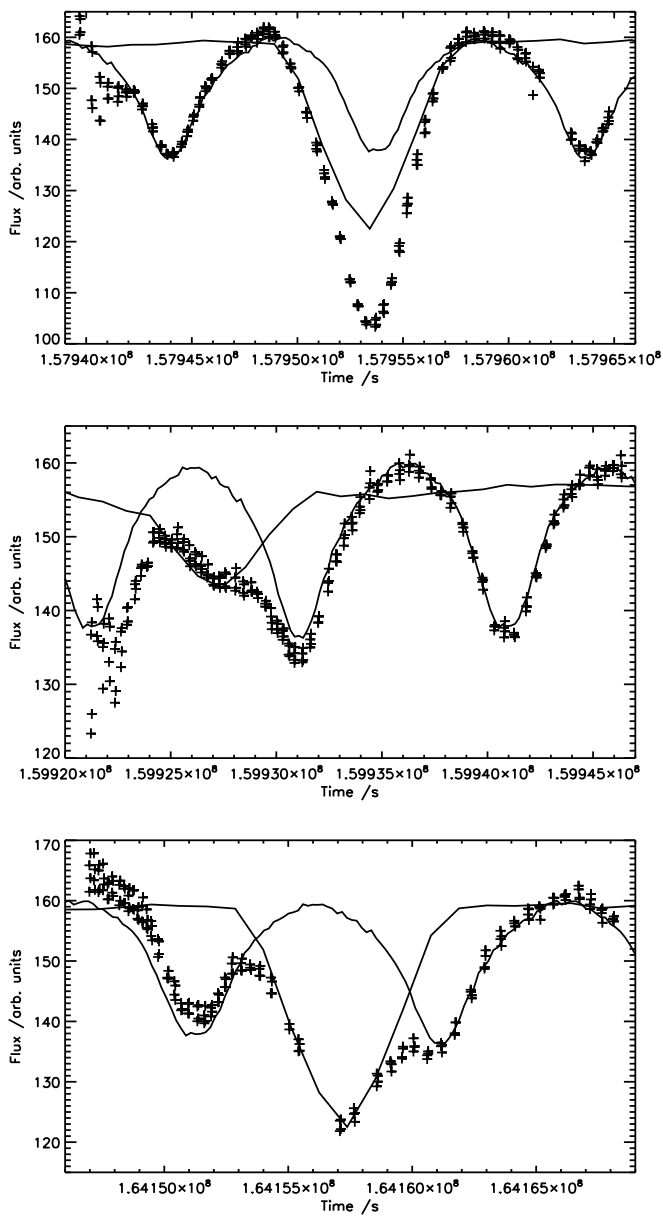}}
\caption{Individual nights of observations of object J093010, with
median binned lightcurves of Algol and contact binaries overplotted
(flux levels adjusted to allow direct comparison).  Top: the Algol's
primary eclipse coincides with the contact system's secondary eclipse.
Middle: the Algol's secondary eclipse nearly coincides with a maximum
of the contact system.  Bottom: the Algol's primary eclipse falls
between the contact system's minima.}
\label{J093010ind}
\end{figure}

Of particular interest is object J093010.  This does not exhibit
significant period change, but its O$-$C diagram showed a surprising
amount of scatter.  In \citet{lohr}, the same object's apparently
significant period change was rejected as the result of contamination
by a nearby star, but here, a fuller explanation was pursued.  Prior
to analysis, the data for J093010 was combined with that from a
`duplicate' object with a similar identifier (see Sect.~\ref{method})
to maximize the available observations.  J093010's lightcurve, folded
at 19\,674.574~s (Fig.~\ref{J093010}, top) then showed a typical contact
binary shape, but with numerous non-random data points below the main
curve.  A visual examination of the object's full lightcurve suggested
the cause was additional deep eclipses on certain individual nights,
implying an additional eclipsing body in the field of view.  A
frequency power spectrum also supported an additional periodic signal
near 1.3~d (Fig.~\ref{powerspec}).  Stripping out the median binned
lightcurve (corresponding to the contact binary) from the data yielded
the lightcurve of an Algol-type eclipsing binary with period
112\,799.109~s (Fig.~\ref{J093010}, middle).  Stripping the median
binned lightcurve of the latter object out of the combined data also
yielded a clean lightcurve for the contact eclipsing binary
(Fig.~\ref{J093010}, bottom).  Figure~\ref{J093010ind} illustrates
how the observations during three representative nights of J093010 are
well-reproduced as the sum of these two eclipsing binary lightcurves.

The question is then, are these two systems physically related, or is
their juxtaposition on the sky coincidental?  Two sources were
observed at this location by Hipparcos as \object{TYC~3807-759-1} and
\object{TYC~3807-759-2}, with equivalent Johnson $V$ magnitudes 9.851 and
10.990 respectively (corresponding to SuperWASP fluxes around 110 and
42), a separation of 1.88$\arcsec$, and a common proper motion (pmRA:
-8.0~mas~yr\textsuperscript{-1}, pmDE:
-9.4~mas~yr\textsuperscript{-1}).  This would seem to strongly favour
an interpretation of the two eclipsing binary systems as being
gravitationally bound in a rare quadruple doubly-eclipsing system.
The distance to TYC~3807-759-1 has been calculated as 35.17~pc,
yielding a separation of 66.1~AU.  Assuming that this separation
corresponds to apastron, and that the total mass for both systems is
$\sim2$ solar masses, a meta-orbital period of $\sim380$ years is
indicated.

Only five other doubly eclipsing quadruple systems have been proposed:
\object{BV~Dra}+\object{BW~Dra}, a contact+contact system
\citep{batten}; \object{V994~Her}, an Algol+Algol system
\citep{lee08}; \object{OGLE-LMC-ECL-16545}, Algol+contact
\citep{graczyk}; \object{KIC~4247791}, Algol+Algol
\citep{lehmann}; and \object{Cze~V343}, another Algol+contact system
\citep{cagas}.  The contact binary in this new quadruple has a shorter
period than any system in the other five quadruples, making it
particularly amenable to further observations.

\subsection{Other systems of note}

Several lightcurves in Fig.~\ref{appfig1} (online only) resemble
$\beta$ Lyrae-type variables (EB), showing notable differences in
primary and secondary eclipse depths: J011732, J022050, J070953,
J093443, J102328, J215826 and J222302 are perhaps the clearest
examples.  Surprisingly for such short-period objects, these may
represent detached or semi-detached systems not in thermal contact.
Indeed, J022050 (=BX~Tri), which shows substantial differences in
depths of minima, as well as a large O'Connell effect
\citep{oconnell}, has been studied in detail by \citet{dimitrov}, who
modelled it as not quite in contact, but with one star nearly filling
its Roche lobe.  Other objects showing pronounced O'Connell effects
(maxima of different heights) are J033242, J084408, J115326, J201816
(if a genuine eclipsing binary), and J221117, of which the last is
particularly striking and unusual in shape.  Finally, we observe
significant variations in average flux in several objects, including
J121206 (which shows distinct well-defined curves at different
levels), J134430, J150957, J151146, J152022, J172717 and J173003
(which show broader banding or continuous variation in average flux
level).  In some cases this variation seems intrinsic to the system,
being observable within a single night's observations with a single
camera; in other cases it probably has instrumental causes e.g. the
object was observed over a long time-base by cameras in both the
northern and southern hemispheres.

\section{Conclusions}

Our new search of the SuperWASP archive yielded 143 plausible
candidate eclipsing binaries with orbital periods $<$20\,000~s, of which
97 are new discoveries.  This updates the findings of \citet{norton},
and provides a useful new sample of extremely short period eclipsing
systems near the cut-off point.  Their period distribution fits neatly
at one end of previous period distributions found for eclipsing
binaries seen by other comparable surveys.  The shape of the
distribution may also inform understanding of the reasons for the
short-period limit for main sequence binaries.

An improved period change detection method was employed with the new
sample, which measured observed times of minima by fitting each night
of observations with the object's own binned mean lightcurve, obtained
from its full data set during period determination.  This approach
confirmed some earlier results in \citet{lohr}, and found significant
evidence for period change in 74 of the candidate binaries, of which
38 are significant at $3\sigma$ or more.  The distribution of
significant period changes found agrees substantially with a previous
comparable distribution, showing equal numbers of systems increasing
and decreasing in period.  Of the highly significant period changes,
none exceed 0.3~s~yr\textsuperscript{-1} in magnitude, and the
half-width at half-maximum of the distribution is
$\sim0.1$~s~yr\textsuperscript{-1}, slightly larger than that found in
a previous study of binaries with longer periods.

Three of the systems found were of particular interest.  J234401,
discussed in \citet{lohr}, continues to show highly significant and
dramatic period changes, though it is currently unclear whether it is
varying in period sinusoidally, or steadily decreasing in period with
unusual rapidity.  J102328 currently shows the most rapid period
decrease of the whole sample: $-0.254\pm
0.037$~s~yr\textsuperscript{-1}, significant at $6\sigma$, which would
indicate a merger timescale $\le$70\,000 years if the decrease
continues.  J093010 appears to be the sixth reported
doubly eclipsing quadruple system, consisting of a contact binary with
a 19\,674.575~s period and an Algol-type binary with a 112\,799.109~s
period, separated by 66.1~AU, and plausibly orbiting each other with a
period of $\sim 400$ years.

\begin{acknowledgements}
The WASP project is funded and operated by Queen's University Belfast,
the Universities of Keele, St. Andrews and Leicester, the Open
University, the Isaac Newton Group, the Instituto de Astrofisica de
Canarias, the South African Astronomical Observatory and by STFC.
This work was supported by the Science and Technology Funding Council
and the Open University.
\end{acknowledgements}

\bibliographystyle{aa}
\bibliography{reflist}

\begin{thebibliography}{33}
\expandafter\ifx\csname natexlab\endcsname\relax\def\natexlab#1{#1}\fi

\bibitem[{Akerlof {et~al.}(2000)Akerlof, Amrose, Balsano, Bloch, Casperson,
  Fletcher, Gisler, Hills, Kehoe, Lee, Marshall, McKay, Pawl, Schaefer,
  Szymanski, \& Wren}]{akerlof}
Akerlof, C., Amrose, S., Balsano, R., {et~al.} 2000, AJ, 119, 1901

\bibitem[{Applegate(1992)}]{applegate}
Applegate, J.~H. 1992, ApJ, 385, 621

\bibitem[{Batten \& Hardie(1965)}]{batten}
Batten, A.~H. \& Hardie, R.~H. 1965, AJ, 70, 666

\bibitem[{Bla\v{z}ko(1907)}]{blazhko}
Bla\v{z}ko, S. 1907, Astron. Nachr., 175, 325

\bibitem[{Caga\v{s} \& Pejcha(2012)}]{cagas}
Caga\v{s}, P. \& Pejcha, O. 2012, A\&A, 544, L3

\bibitem[{Christopoulou {et~al.}(2012)Christopoulou, Papageorgiou, Vasileiadis,
  \& Tsantilas}]{christopoulou}
Christopoulou, P.~E., Papageorgiou, A., Vasileiadis, T., \& Tsantilas, S. 2012,
  AJ, 144, 149

\bibitem[{Dimitrov \& Kjurkchieva(2010)}]{dimitrov}
Dimitrov, D.~P. \& Kjurkchieva, D.~P. 2010, MNRAS, 406, 2559

\bibitem[{Graczyk {et~al.}(2011)Graczyk, Soszy\'{n}ski, Poleski, Pietrzyński,
  Udalski, Szymański, Kubiak, Wyrzykowski, \& Ulaczyk}]{graczyk}
Graczyk, D., Soszy\'{n}ski, I., Poleski, R., {et~al.} 2011, Acta Astron., 61,
  103

\bibitem[{Hilditch(2001)}]{hilditch}
Hilditch, R.~W. 2001, An Introduction to Close Binary Stars (Cambridge, UK:
  Cambridge University Press)

\bibitem[{Jiang {et~al.}(2012)Jiang, Han, Ge, Yang, \& Li}]{jiang}
Jiang, D., Han, Z., Ge, H., Yang, L., \& Li, L. 2012, MNRAS, 421, 2769

\bibitem[{Kubiak {et~al.}(2006)Kubiak, Udalski, \& Szyma\'{n}ski}]{kubiak}
Kubiak, M., Udalski, A., \& Szyma\'{n}ski, M.~K. 2006, Acta Astron., 56, 253

\bibitem[{Kwee \& van Woerden(1956)}]{kwee}
Kwee, K.~K. \& van Woerden, H. 1956, Bulletin of the Astronomical Institutes of
  the Netherlands, 12, 327

\bibitem[{Lee {et~al.}(2008)Lee, Kim, Lee, Kim, Jeon, Kim, Yoon, \&
  Humphrey}]{lee08}
Lee, C.~U., Kim, S.~L., Lee, J.~W., {et~al.} 2008, MNRAS, 389, 1630

\bibitem[{Lee {et~al.}(2011)Lee, Lee, Kim, Kim, \& Park}]{lee}
Lee, J.~W., Lee, C.-U., Kim, S.-L., Kim, H.-I., \& Park, J.-H. 2011, PASP, 123,
  34

\bibitem[{Lehmann {et~al.}(2012)Lehmann, Zechmeister, Dreizler, Schuh, \&
  Kanzler}]{lehmann}
Lehmann, H., Zechmeister, M., Dreizler, S., Schuh, S., \& Kanzler, R. 2012,
  A\&A, 541, A105

\bibitem[{Levenberg(1944)}]{lev}
Levenberg, K. 1944, Quarterly of Applied Mathematics, 2, 164

\bibitem[{Lohr {et~al.}(2012)Lohr, Norton, Kolb, Anderson, Faedi, \&
  West}]{lohr}
Lohr, M.~E., Norton, A.~J., Kolb, U.~C., {et~al.} 2012, A\&A, 542, A124

\bibitem[{Marquardt(1963)}]{marq}
Marquardt, D.~W. 1963, SIAM Journal on Applied Mathematics, 11, 431

\bibitem[{Mazeh {et~al.}(2006)Mazeh, Tamuz, Zucker, Udalski, Consortium,
  Arnold, Bouchy, \& Moutou}]{mazeh}
Mazeh, T., Tamuz, O., Zucker, S., {et~al.} 2006, in Tenth Anniversary of 51
  Peg-b: Status of and prospects for hot Jupiter studies. Colloquium held at
  Observatoire de Haute Provence, France, August 22-25, 2005, ed. L.~Arnold,
  F.~Bouchy, \& C.~Moutou (Paris: Frontier Group), 165--172

\bibitem[{Nefs {et~al.}(2012)Nefs, Birkby, Snellen, Hodgkin, Pinfield, Sipocz,
  Kovacs, Mislis, Saglia, Koppenhoefer, Cruz, Barrado, Martin, Goulding, Stoev,
  Zendejas, del Burgo, Cappetta, \& Pavlenko}]{nefs}
Nefs, S.~V., Birkby, J.~L., Snellen, I. A.~G., {et~al.} 2012, MNRAS, 425, 950

\bibitem[{Norton {et~al.}(2011)Norton, Payne, Evans, West, Wheatley, Anderson,
  Barros, Butters, Cameron, Christian, Enoch, Faedi, Haswell, Hellier, Holmes,
  Horne, Kane, Lister, Maxted, Parley, Pollacco, Simpson, Skillen, Smalley,
  Southworth, \& Street}]{norton}
Norton, A.~J., Payne, S.~G., Evans, T., {et~al.} 2011, A\&A, 528, A90

\bibitem[{Norton {et~al.}(2007)Norton, Wheatley, West, Haswell, Street,
  Cameron, Christian, Clarkson, Enoch, Gallaway, Hellier, Horne, Irwin, Kane,
  Lister, Nicholas, Parley, Pollacco, Ryans, Skillen, \& Wilson}]{norton07}
Norton, A.~J., Wheatley, P.~J., West, R.~G., {et~al.} 2007, A\&A, 467, 785

\bibitem[{O'Connell(1951)}]{oconnell}
O'Connell, D. J.~K. 1951, Publications of the Riverview College Observatory, 2,
  85

\bibitem[{Paczy\'{n}ski {et~al.}(2006)Paczy\'{n}ski, Szczygie\l, Pilecki, \&
  Pojma\'{n}ski}]{paczynski}
Paczy\'{n}ski, B., Szczygie\l, D.~M., Pilecki, B., \& Pojma\'{n}ski, G. 2006,
  MNRAS, 368, 1311

\bibitem[{Pollacco {et~al.}(2006)Pollacco, Skillen, Cameron, Christian,
  Hellier, Irwin, Lister, Street, West, Anderson, Clarkson, Deeg, Enoch, Evans,
  Fitzsimmons, Haswell, Hodgkin, Horne, Kane, Keenan, Maxted, Norton, Osborne,
  Parley, Ryans, Smalley, Wheatley, \& Wilson}]{pollacco}
Pollacco, D.~L., Skillen, I., Cameron, A.~C., {et~al.} 2006, PASP, 118, 1407

\bibitem[{Pribulla {et~al.}(2008)Pribulla, Balud'ansk\'{y}, Dubovsk\'{y},
  Kudzej, Parimucha, Siwak, \& Vaňko}]{pribulla08}
Pribulla, T., Balud'ansk\'{y}, D., Dubovsk\'{y}, P., {et~al.} 2008, MNRAS, 390,
  798

\bibitem[{Pribulla {et~al.}(2012)Pribulla, Va\v{n}ko, von Eiff, Andreev,
  Aslant\"{u}rk, Awadalla, Baluďanský, Bonanno, Božić, Catanzaro, Çelik,
  Christopoulou, Covino, Cusano, Dimitrov, Dubovský, Esmer, Frasca, Hambálek,
  Hanna, Hanslmeier, Kalomeni, Kjurkchieva, Krushevska, Kudzej, Kundra,
  Kuznyetsova, Lee, Leitzinger, Maciejewski, Moldovan, Morais, Mugrauer,
  Neuhäuser, Niedzielski, Odert, Ohlert, Özavcı, Papageorgiou, Parimucha,
  Poddaný, Pop, Raetz, Raetz, Romanyuk, Ruždjak, Schulz, Şenavcı, Szalai,
  Székely, Sudar, Tezcan, Törün, Turcu, Vince, \& Zejda}]{pribulla}
Pribulla, T., Va\v{n}ko, M., von Eiff, M.~A., {et~al.} 2012, Astron. Nachr.,
  333, 754

\bibitem[{Rucinski(1992)}]{ruc92}
Rucinski, S.~M. 1992, AJ, 103, 960

\bibitem[{Rucinski(2007)}]{ruc07}
Rucinski, S.~M. 2007, MNRAS, 382, 393

\bibitem[{Stepie\'{n}(2006)}]{stepien}
Stepie\'{n}, K. 2006, Acta Astron., 56, 347

\bibitem[{Stepie\'{n} \& Gazeas(2012)}]{stepien12}
Stepie\'{n}, K. \& Gazeas, K. 2012, Acta Astron., 62, 153

\bibitem[{Szyma\'{n}ski {et~al.}(2001)Szyma\'{n}ski, Kubiak, \&
  Udalski}]{szymanski}
Szyma\'{n}ski, M., Kubiak, M., \& Udalski, A. 2001, Acta Astron., 51, 259

\bibitem[{Tamuz {et~al.}(2005)Tamuz, Mazeh, \& Zucker}]{tamuz}
Tamuz, O., Mazeh, T., \& Zucker, S. 2005, MNRAS, 356, 1466

\end{thebibliography}

\begin{longtable}{l l r r r r r r r}
\caption{\label{143data} Period and period change determinations for 143 eclipsing binary candidates.  Only period changes significant at 1~$\sigma$ are shown.}\\
\hline\hline
SuperWASP ID (1SWASP & Previous ID as & Mean SW & Prim. & Sec. & $P$ & $\dot{P}$ & $\delta \dot{P}$ & Signif.\\
Jhhmmss.ss$\pm$ddmmss.s) & variable star & V mag & depth & depth & ($\pm$0.0005~s) & (s~yr\textsuperscript{-1}) & (s~yr\textsuperscript{-1}) & ($\sigma$)\\
\hline
\endfirsthead
\caption{continued.}\\
\hline\hline
SuperWASP ID (1SWASP & Previous ID as & Mean SW & Prim. & Sec. & $P$ & $\dot{P}$ & $\delta \dot{P}$ & Signif.\\
Jhhmmss.ss$\pm$ddmmss.s) & variable star & V mag & depth & depth & ($\pm$0.0005~s) & (s~yr\textsuperscript{-1}) & (s~yr\textsuperscript{-1}) & ($\sigma$)\\
\hline
\endhead
J000205.32+381321.5 & & 15.16 & 15.33 & 15.27 & 18064.547 & 0.036 & 0.026 & 1 \\
J003033.05+574347.6 & 27\footnote{Numbers refer to Table 1 in \citet{norton}; duplicate objects are in parentheses.} & 15.13 & 15.63 & 15.57 & 19579.837 & & & \\
J004050.63+071613.9 & 15 & 12.30 & 12.37 & 12.37 & 19809.222 & & & \\
J004545.23$-$244516.2 & & 15.37 & 15.52 & 15.52 & 19037.467 & 0.037 & 0.020 & 1 \\
J010340.37$-$172138.8 & & 14.80 & 14.99 & 14.96 & 19719.757 & 0.029 & 0.012 & 2 \\
J010642.20$-$330857.9 & & 13.95 & 14.14 & 14.09 & 19187.558 & 0.0549 & 0.0073 & 7 \\
J011732.10+525204.9 & & 13.98 & 14.09 & 14.02 & 19350.996 & & & \\
J015100.23$-$100524.2 & & 14.73 & 15.15 & 15.15 & 18532.806 & 0.072 & 0.021 & 3 \\
J022050.85+332047.6 & 53; BX~Tri & 13.25 & 13.44 & 13.35 & 16643.640 & $-$0.0290 & 0.0074 & 3 \\
J022727.03+115641.7 & 47 & 15.08 & 15.23 & 15.22 & 18226.385 & 0.131 & 0.076 & 1 \\
J023459.08$-$393704.6 & & 15.69 & 16.27 & 16.24 & 19186.467 & & & \\
J024148.62+372848.3 & & 15.33 & 15.54 & 15.52 & 18986.466 & & & \\
J025054.80+012357.5 & & 12.03 & 12.05 & 12.05 & 19929.602 & $-$0.088 & 0.033 & 2 \\
J025959.18$-$395812.3 & & 14.50 & 14.67 & 14.63 & 19254.539 & $-$0.065 & 0.022 & 2 \\
J030749.87$-$365201.7 & 26 & 15.14 & 15.57 & 15.54 & 19584.393 & & & \\
J031700.67+190839.6 & & 12.63 & 12.68 & 12.67 & 19496.300 & & & \\
J033242.81$-$085525.9 & LL~Eri\footnote{Identified as ellipsoidal variable in GCVS.} & 13.31 & 13.36 & 13.34 & 17420.557 & 0.83 & 0.32 & 2 \\
J034439.97+030425.5 & 11 & 14.33 & 14.46 & 14.42 & 19861.387 & & & \\
J040615.79$-$425002.3 & 34 & 14.31 & 14.68 & 14.63 & 19209.987 & $-$0.0294 & 0.0047 & 6 \\
J041120.40$-$230232.3 & 43 & 14.09 & 14.38 & 14.30 & 18690.347 & $-$0.124 & 0.098 & 1 \\
J041655.13$-$492709.8 & 5 & 15.14 & 15.40 & 15.31 & 19959.816 & & & \\
J042200.64$-$450312.5 & & 13.63 & 13.69 & 13.69 & 18843.443 & 0.0657 & 0.0078 & 8 \\
J044132.96+440613.7 & 19 & 14.38 & 14.59 & 14.57 & 19712.589 & & & \\
J050128.17$-$041206.9 & & 12.69 & 12.72 & 12.72 & 19512.511 & 0.070 & 0.035 & 2 \\
J050520.94$-$374338.6 & & 16.34 & 16.86 & 16.69 & 19933.344 & 0.150 & 0.032 & 4 \\
J050723.00$-$502512.9 & & 14.83 & 15.02 & 15.00 & 18251.638 & & & \\
J050904.45$-$074144.4 & 13 & 13.51 & 13.95 & 13.90 & 19835.305 & & & \\
J051459.80$-$021923.6 & (6) & 13.76 & 13.85 & 13.83 & 19950.123 & $-$0.101 & 0.034 & 2 \\
J052036.84+030402.1 & 1 & 12.46 & 12.62 & 12.62 & 19993.297 & $-$0.025 & 0.011 & 2 \\
J052825.85+093943.7 & & 14.06 & 14.15 & 14.13 & 19068.472 & & & \\
J052926.88+461147.5 & & 14.66 & 14.91 & 14.86 & 19581.884 & & & \\
J055215.51$-$551950.8 & & 15.50 & 15.65 & 15.61 & 19637.229 & & & \\
J055416.98+442534.0 & (39) & 12.73 & 12.95 & 12.93 & 18877.974 & & & \\
J060334.52$-$283427.1 & & 14.78 & 14.86 & 14.86 & 17828.542 & & & \\
J061011.73$-$345809.0 & & 11.6 & 11.61 & 11.61 & 19884.225 & & & \\
J061236.44$-$281553.0 & & 15.65 & 15.77 & 15.75 & 19955.086 & 0.114 & 0.036 & 3 \\
J061850.43+220511.9 & & 13.87 & 14.05 & 14.01 & 18523.570 & & & \\
J062634.80$-$385650.1 & & 15.69 & 15.87 & 15.80 & 19326.892 & 0.152 & 0.019 & 8 \\
J070953.45+364417.3 & & 12.06 & 12.10 & 12.08 & 19271.205 & 0.66 & 0.30 & 2 \\
J072504.73+410212.3 & & 9.75 & 9.76 & 9.76 & 19479.807 & & & \\
J074658.62+224448.5 & 38 & 14.27 & 14.72 & 14.59 & 19081.403 & & & \\
J075102.16+342405.3 & & 14.69 & 14.91 & 14.88 & 18072.478 & & & \\
J075149.14+362250.9 & & 11.28 & 11.30 & 11.30 & 19974.770 & & & \\
J080150.03+471433.8 & 42 & 13.62 & 14.05 & 14.04 & 18793.174 & $-$0.066 & 0.043 & 1 \\
J084408.68$-$040640.1 & & 16.02 & 17.06 & 16.55 & 18812.053 & & & \\
J084925.17$-$151516.5 & & 15.27 & 15.37 & 15.37 & 17276.788 & $-$0.164 & 0.058 & 2 \\
J090758.16$-$153811.8 & & 13.53 & 13.56 & 13.55 & 19775.555 & & & \\
J092339.29$-$412648.9 & & 9.71 & 9.72 & 9.72 & 17533.446 & & & \\
J092754.99$-$391053.4 & (30) & 12.01 & 12.15 & 12.12 & 19469.694 & $-$0.019 & 0.014 & 1 \\
J093010.78+533859.5 & (22) & 9.56 & 9.67 & 9.66 & 19674.598 & & & \\
J093443.60+420831.9 & & 13.78 & 13.96 & 13.82 & 19201.572 & $-$0.095 & 0.023 & 4 \\
J095706.80$-$201408.7 & & 15.55 & 15.82 & 15.77 & 19663.870 & & & \\
J101618.12$-$085531.0 & & 14.14 & 14.26 & 14.24 & 18546.804 & 0.065 & 0.020 & 3 \\
J102328.57$-$153951.7 & & 14.13 & 14.23 & 14.17 & 18125.146 & $-$0.254 & 0.037 & 6 \\
J104125.56$-$145842.3 & & 15.47 & 15.67 & 15.65 & 19501.887 & 0.260 & 0.036 & 7 \\
J104942.44+141021.5 & & 14.93 & 15.09 & 15.06 & 19854.610 & & & \\
J105455.18$-$352052.5 & & 12.18 & 12.23 & 12.22 & 19981.603 & 0.0507 & 0.0083 & 6 \\
J111501.66$-$361254.2 & & 15.45 & 15.67 & 15.61 & 19008.433 & 0.0933 & 0.0090 & 10 \\
J111931.48$-$395048.2 & 14; ASAS & 10.8 & 10.87 & 10.87 & 19827.728 & 0.0216 & 0.0022 & 9 \\
 & J111932$-$3950.8 & & & & & & & \\
J114929.22$-$423049.0 & 23 & 14.44 & 14.99 & 14.83 & 19639.528 & $-$0.0201 & 0.0032 & 6 \\
J115326.51+060756.0 & & 14.69 & 14.83 & 14.80 & 19754.467 & 0.049 & 0.024 & 2 \\
J115557.62+072009.1 & (25) & 15.36 & 15.67 & 15.65 & 19614.248 & & & \\
J115605.88$-$091300.5 & 48 & 12.49 & 12.52 & 12.52 & 18222.626 & $-$0.045 & 0.018 & 2 \\
J120110.98$-$220210.8 & 24 & 14.27 & 14.36 & 14.33 & 19627.826 & $-$0.253 & 0.012 & 20 \\
J120230.52$-$211314.4 & & 16.31 & 16.64 & 16.64 & 19998.711 & & & \\
J121206.02+223158.7 & CC~Com & 11.61 & 12.20 & 12.08 & 19067.304 & 0.0265 & 0.0011 & 23 \\
J121359.79$-$414742.7 & & 13.48 & 13.51 & 13.51 & 18944.739 & 0.0348 & 0.0077 & 4 \\
J121906.35$-$240056.9 & 29 & 15.31 & 15.73 & 15.55 & 19558.142 & $-$0.0771 & 0.0074 & 10 \\
J123148.12$-$020602.3 & & 14.97 & 15.23 & 15.19 & 19578.952 & 0.061 & 0.027 & 2 \\
J130111.22+420214.0 & & 15.34 & 15.69 & 15.55 & 19477.594 & 0.0174 & 0.0078 & 2 \\
J130711.90+084159.9 & & 15.23 & 15.37 & 15.36 & 19679.485 & & & \\
J130920.49$-$340919.9 & 33 & 14.06 & 14.18 & 14.17 & 19253.458 & $-$0.0276 & 0.0091 & 3 \\
J132308.74+424613.3 & & 14.06 & 14.22 & 14.18 & 19451.259 & $-$0.24 & 0.13 & 1 \\
J133105.91+121538.0 & 41 & 10.46 & 10.59 & 10.55 & 18836.380 & $-$0.0202 & 0.0038 & 5 \\
J133417.80+394314.4 & & 14.86 & 15.03 & 15.03 & 19775.388 & & & \\
J134430.51$-$270302.8 & & 10.34 & 10.37 & 10.36 & 19841.554 & $-$0.130 & 0.025 & 5 \\
J135403.76$-$462948.7 & & 14.46 & 14.64 & 14.62 & 19762.315 & 0.022 & 0.015 & 1 \\
J140533.33+114639.1 & & 16.20 & 16.51 & 16.42 & 19450.626 & 0.156 & 0.062 & 2 \\
J142312.63$-$222425.1 & 51 & 13.21 & 13.24 & 13.24 & 18112.855 & $-$0.1152 & 0.0099 & 11 \\
J144331.57$-$421626.8 & & 14.60 & 14.80 & 14.75 & 18598.876 & $-$0.0116 & 0.0072 & 1 \\
J150957.56$-$115308.4 & & 14.80 & 15.09 & 15.07 & 19787.358 & & & \\
J151144.56+165426.4 & & 15.03 & 15.43 & 15.43 & 18996.306 & $-$0.0100 & 0.0075 & 1 \\
J151146.20$-$354721.9 & & 12.14 & 12.20 & 12.19 & 19227.102 & $-$0.0167 & 0.0096 & 1 \\
J151652.90+004835.8 & 49 & 14.01 & 14.08 & 14.07 & 18207.339 & & & \\
J152022.78$-$340512.8 & & 15.16 & 15.42 & 15.40 & 18893.115 & & & \\
J153951.12+105420.7 & & 15.36 & 15.97 & 15.82 & 19070.320 & $-$0.041 & 0.028 & 1 \\
J160156.04+202821.6 & 28 & 14.31 & 14.80 & 14.72 & 19572.136 & 0.104 & 0.011 & 9 \\
J160202.07+121213.5 & & 13.27 & 13.34 & 13.32 & 18973.486 & 0.032 & 0.023 & 1 \\
J161335.80$-$284722.2 & (12) & 12.55 & 12.94 & 12.83 & 19852.821 & $-$0.158 & 0.018 & 8 \\
J161858.05+261303.5 & & 13.21 & 13.24 & 13.23 & 19766.679 & & & \\
J162117.36+441254.2 & & 15.13 & 15.36 & 15.26 & 17958.417 & 0.0084 & 0.0060 & 1 \\
J162841.41$-$334419.8 & & 13.96 & 14.02 & 13.99 & 17598.403 & 0.203 & 0.063 & 3 \\
J164349.61+325637.8 & ROTSE1~J164349.58 & 12.12 & 12.23 & 12.21 & 19447.858 & 0.0024 & 0.0008 & 2 \\
 & +325637.8\footnote{Identified as $\delta$~Scuti variable by ROTSE.} & & & & & & & \\
J165734.53+274827.7 & & 14.51 & 14.63 & 14.62 & 19396.507 & & & \\
J172717.97+431624.0 & & 13.43 & 13.5 & 13.49 & 19446.039 & $-$0.0035 & 0.0022 & 1 \\
J173003.21+344509.4 & 32 & 13.94 & 14.13 & 14.09 & 19328.926 & 0.0169 & 0.0015 & 11 \\
J180947.64+490255.0 & 20; V1104~Her & 14.13 & 14.85 & 14.63 & 19688.482 & 0.0101 & 0.0008 & 12 \\
J183738.17+402427.2 & 36 & 14.56 & 14.81 & 14.78 & 19121.225 & 0.0039 & 0.0023 & 1 \\
J192218.68$-$303313.1 & & 12.31 & 12.36 & 12.35 & 18098.762 & $-$0.0876 & 0.0089 & 9 \\
J193127.17+465809.1 & & 12.12 & 12.14 & 12.14 & 17075.186 & & & \\
J193537.06$-$401409.1 & & 15.92 & 16.24 & 16.14 & 19517.575 & & & \\
J194726.58$-$243941.0 & & 12.42 & 12.45 & 12.45 & 17568.849 & & & \\
J195730.89+000705.1 & & 12.64 & 12.69 & 12.67 & 19563.661 & & & \\
J200059.78+054408.9 & & 14.31 & 14.44 & 14.43 & 17771.663 & & & \\
J200503.05$-$343726.5 & & 15.71 & 16.34 & 16.17 & 19775.546 & $-$0.021 & 0.011 & 1 \\
J200756.54$-$163408.0 & & 11.68 & 11.73 & 11.72 & 19226.539 & 0.1012 & 0.0066 & 15 \\
J201208.72+083509.8 & & 12.55 & 12.58 & 12.57 & 18642.212 & & & \\
J201808.68$-$231443.0 & & 14.41 & 14.55 & 14.51 & 19683.179 & $-$0.025 & 0.012 & 2 \\
J201816.85+112452.8 & & 10.34 & 10.35 & 10.35 & 16101.400 & & & \\
J203831.39$-$593324.1 & & 15.23 & 15.34 & 15.34 & 18810.617 & $-$0.084 & 0.029 & 2 \\
J204843.90$-$350912.7 & & 13.29 & 13.34 & 13.32 & 19770.864 & & & \\
J204932.94$-$654025.8 & & 14.22 & 14.31 & 14.27 & 19864.254 & & & \\
J210318.76+021002.2 & 16 & 13.02 & 13.04 & 13.04 & 19750.171 & $-$0.0672 & 0.0098 & 6 \\
J210423.94+073104.8 & (52) & 13.67 & 13.72 & 13.72 & 18065.088 & 0.0332 & 0.0039 & 8 \\
J211359.46+122712.4 & & 14.87 & 15.08 & 15.02 & 19190.122 & $-$0.0508 & 0.0035 & 14 \\
J211625.31+251755.4 & & 12.95 & 13.01 & 12.99 & 18782.483 & & & \\
J212009.70$-$185220.8 & & 14.19 & 14.29 & 14.27 & 18817.674 & & & \\
J212454.61+203030.8 & 21 & 14.55 & 14.75 & 14.72 & 19684.798 & & & \\
J212808.86+151622.0 & 31 & 14.66 & 15.01 & 14.89 & 19426.320 & & & \\
J212813.35$-$520029.1 & & 15.63 & 15.77 & 15.75 & 19913.352 & & & \\
J213252.93$-$441822.6 & & 16.69 & 17.22 & 17.13 & 19114.669 & & & \\
J214046.44+130716.6 & & 12.50 & 12.53 & 12.53 & 19523.322 & & & \\
J214510.25$-$494401.1 & 18 & 14.69 & 14.79 & 14.78 & 19712.977 & & & \\
J215826.52+253437.4 & & 13.14 & 13.19 & 13.14 & 19233.919 & & & \\
J220235.74+311909.7 & & 14.02 & 14.20 & 14.19 & 19049.179 & 0.068 & 0.017 & 3 \\
J220524.25+204151.1 & & 15.34 & 15.55 & 15.50 & 19922.164 & & & \\
J220734.47+265528.6 & 2 & 14.34 & 14.52 & 14.50 & 19978.747 & & & \\
J221045.95+211842.3 & & 15.48 & 15.71 & 15.71 & 19745.719 & & & \\
J221058.82+251123.4 & 45 & 15.23 & 15.37 & 15.34 & 18402.959 & 0.073 & 0.035 & 2 \\
J221117.26$-$150216.6 & & 13.73 & 13.92 & 13.87 & 18597.484 & & & \\
J222302.02+195031.8 & & 15.75 & 16.05 & 15.89 & 19455.256 & & & \\
J222514.69+361643.0 & & 12.63 & 12.65 & 12.65 & 19416.976 & & & \\
J224747.20$-$351849.3 & 40 & 13.97 & 14.10 & 14.05 & 18853.816 & 0.081 & 0.010 & 8 \\
J230042.26$-$365859.7 & & 15.47 & 15.67 & 15.63 & 19448.414 & $-$0.029 & 0.020 & 1 \\
J231505.30$-$010617.0 & & 13.47 & 13.57 & 13.56 & 19836.479 & & & \\
J231839.72+352848.2 & & 14.91 & 15.02 & 14.98 & 17388.474 & & & \\
J231943.31+134121.4 & & 14.34 & 14.44 & 14.42 & 19944.234 & & & \\
J232610.13$-$294146.6 & (8) & 13.58 & 13.92 & 13.84 & 19882.087 & & & \\
J233120.96$-$145814.2 & & 13.85 & 13.99 & 13.97 & 18657.047 & & & \\
J234401.81$-$212229.1 & 44 & 11.68 & 11.75 & 11.73 & 18461.639 & $-$0.1422 & 0.0041 & 35 \\
J235216.40+465044.9 & & 10.41 & 10.42 & 10.42 & 18416.616 & & & \\
J235333.60+455245.8 & 7 & 14.88 & 15.10 & 15.06 & 19935.681 & & & \\
J235935.22+362001.5 & & 11.76 & 11.77 & 11.77 & 17424.406 & $-$0.120 & 0.023 & 5 \\
\end{longtable}

\begin{figure*}
\centering
\includegraphics[width=16cm]{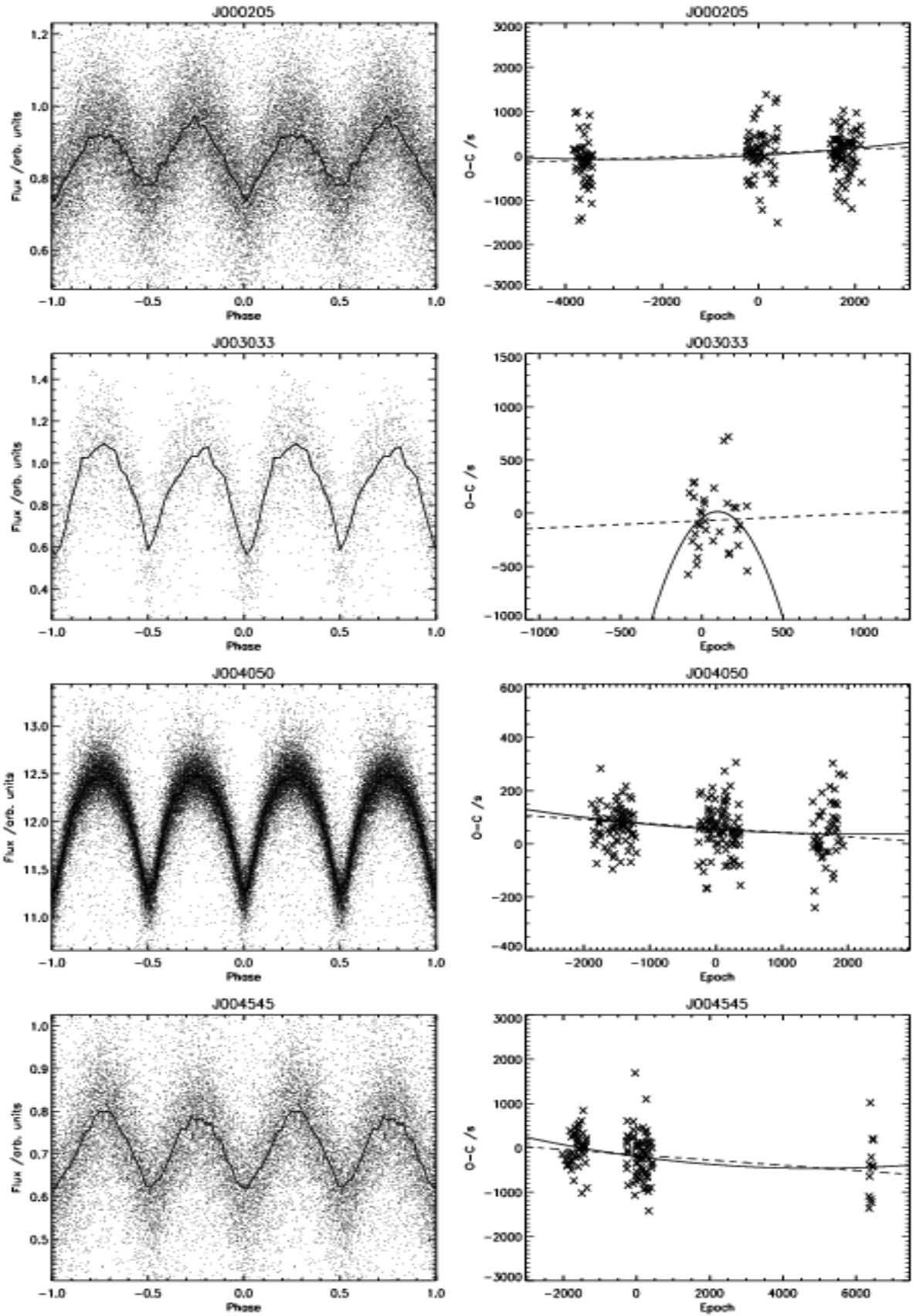}
\caption{Folded lightcurves (left) and O$-$C diagrams (right) for 143 objects.  On the O$-$C diagrams, the dotted lines represent the best linear fit to the data, and the solid lines the best quadratic fit.  Note that many O$-$C diagrams represent non-significant period change, or were disregarded due to paucity of data: see Table~\ref{143data} for significant results.}
\label{appfig1}
\end{figure*}

\clearpage

\begin{center}
\includegraphics[width=16cm]{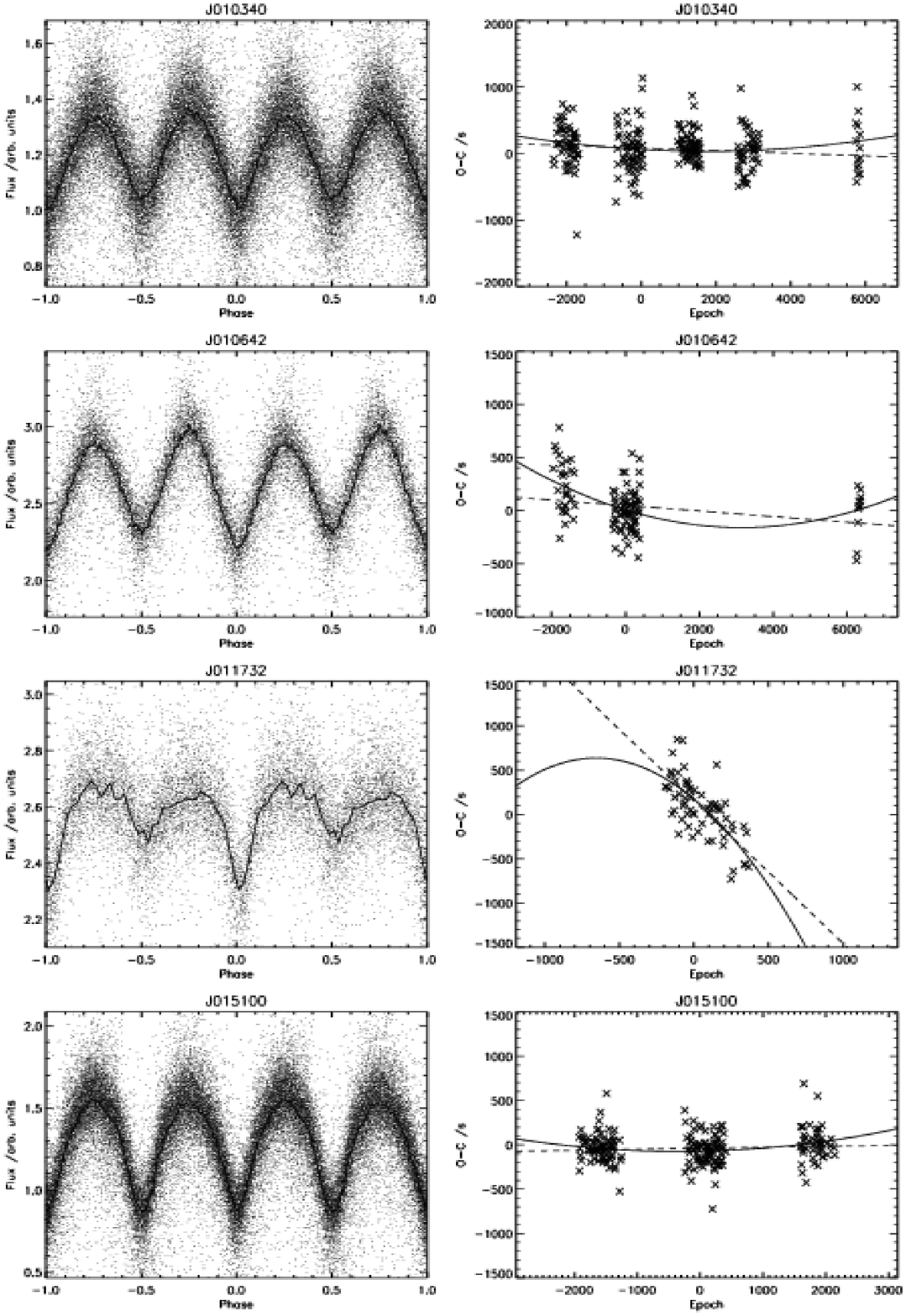}
\end{center}\textbf{Fig.~\ref{appfig1}} continued. \clearpage

\begin{center} \includegraphics[width=16cm]{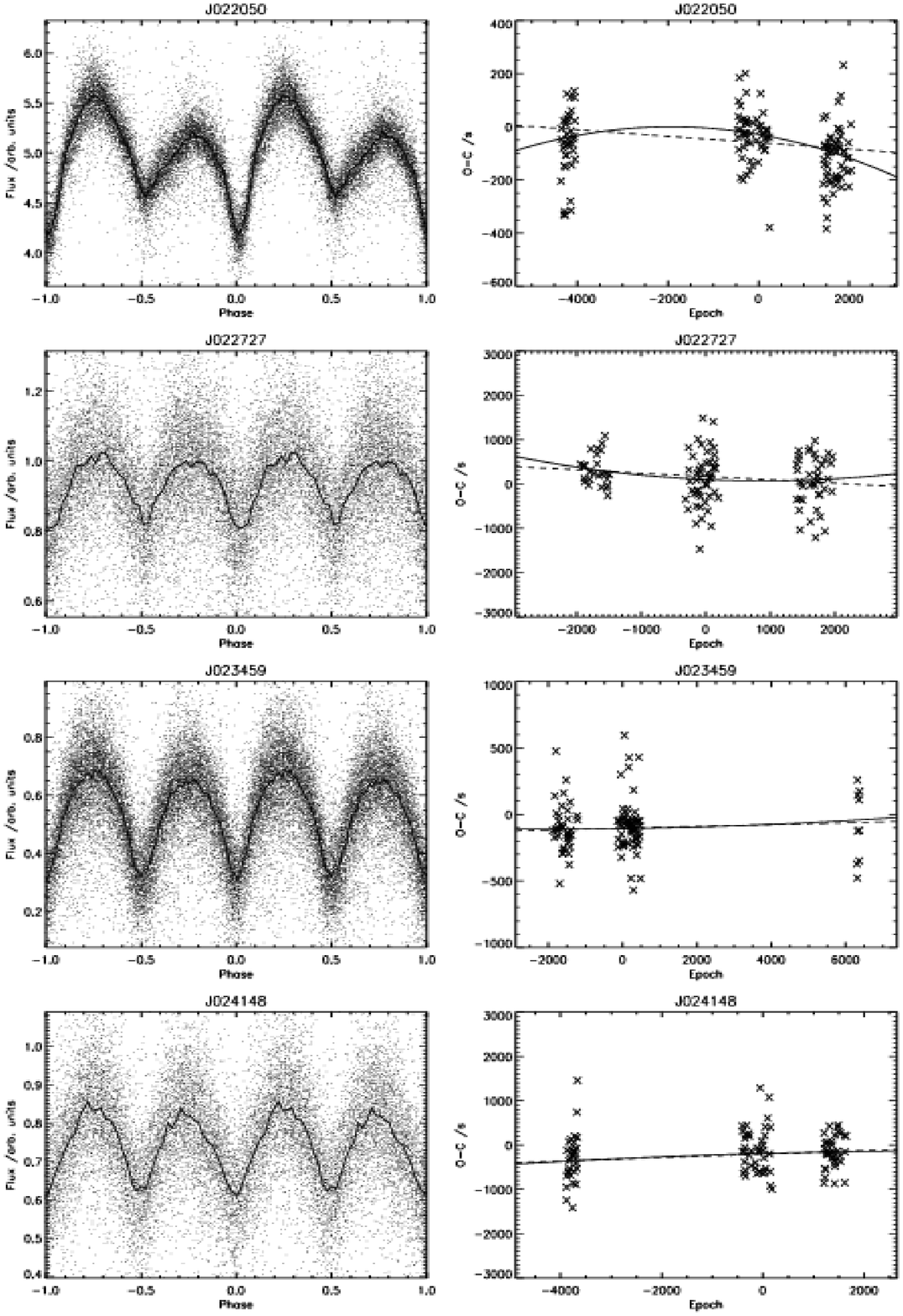}
\end{center}\textbf{Fig.~\ref{appfig1}} continued. \clearpage

\begin{center} \includegraphics[width=16cm]{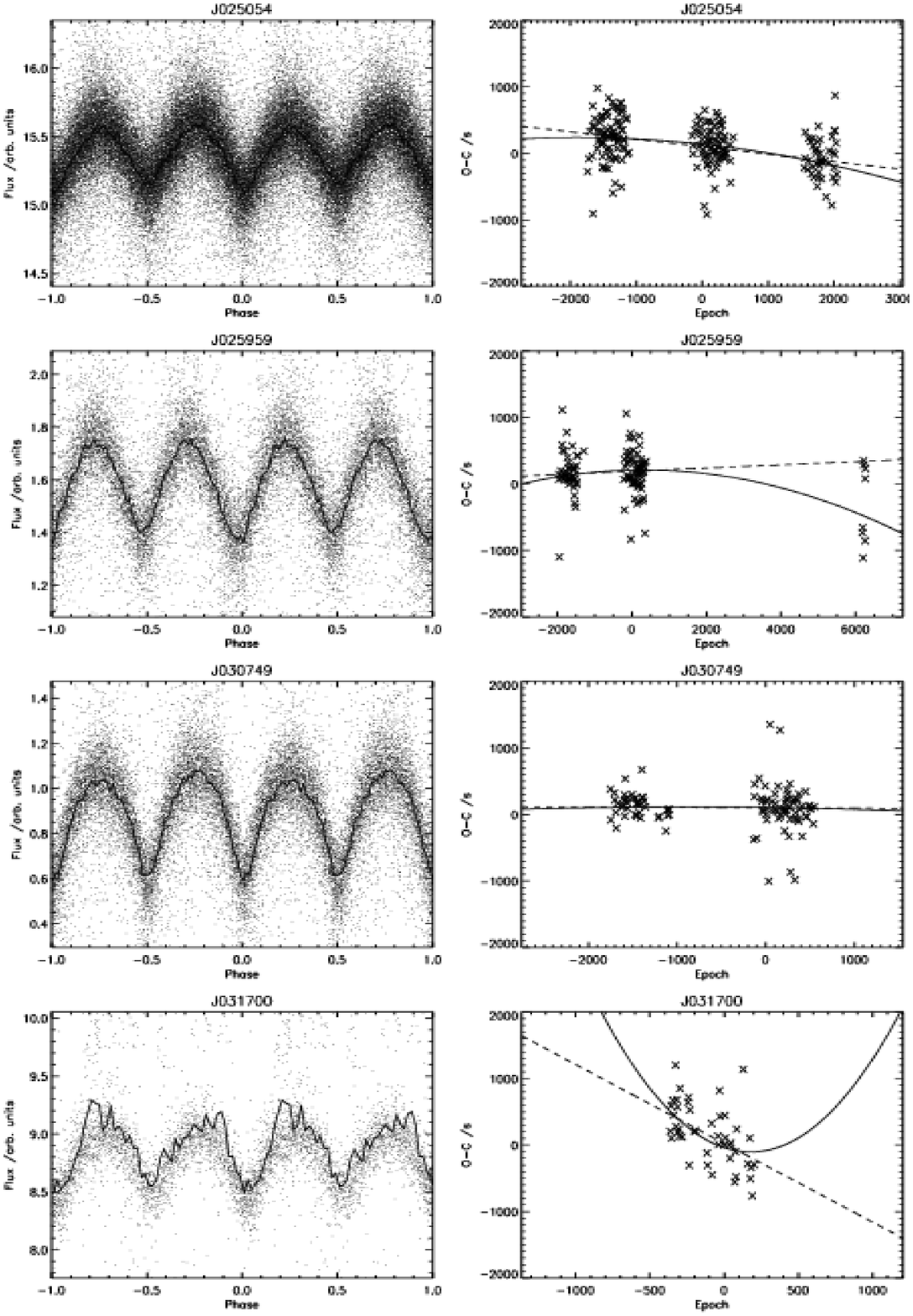}
\end{center}\textbf{Fig.~\ref{appfig1}} continued. \clearpage

\begin{center} \includegraphics[width=16cm]{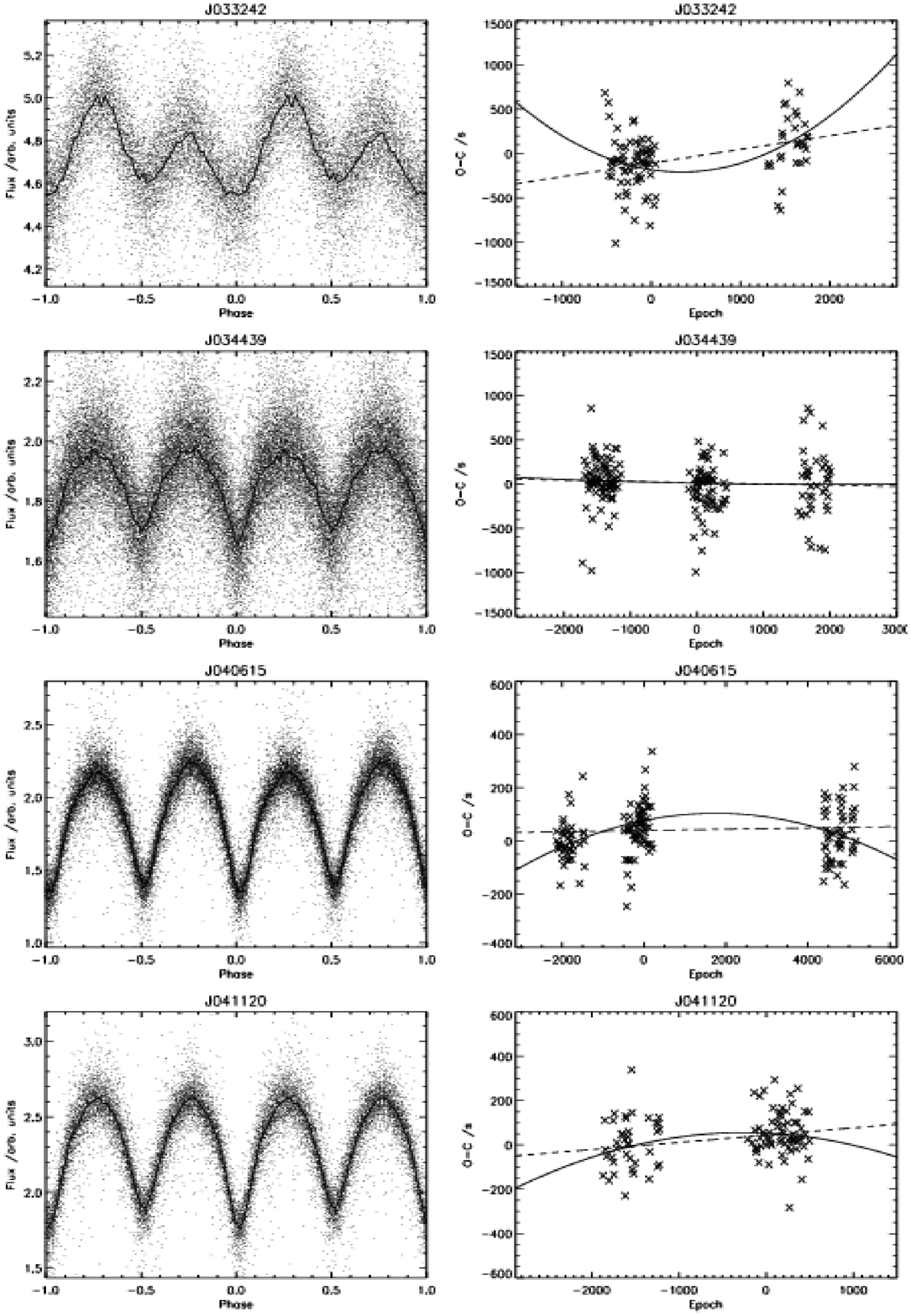}
\end{center}\textbf{Fig.~\ref{appfig1}} continued. \clearpage

\begin{center} \includegraphics[width=16cm]{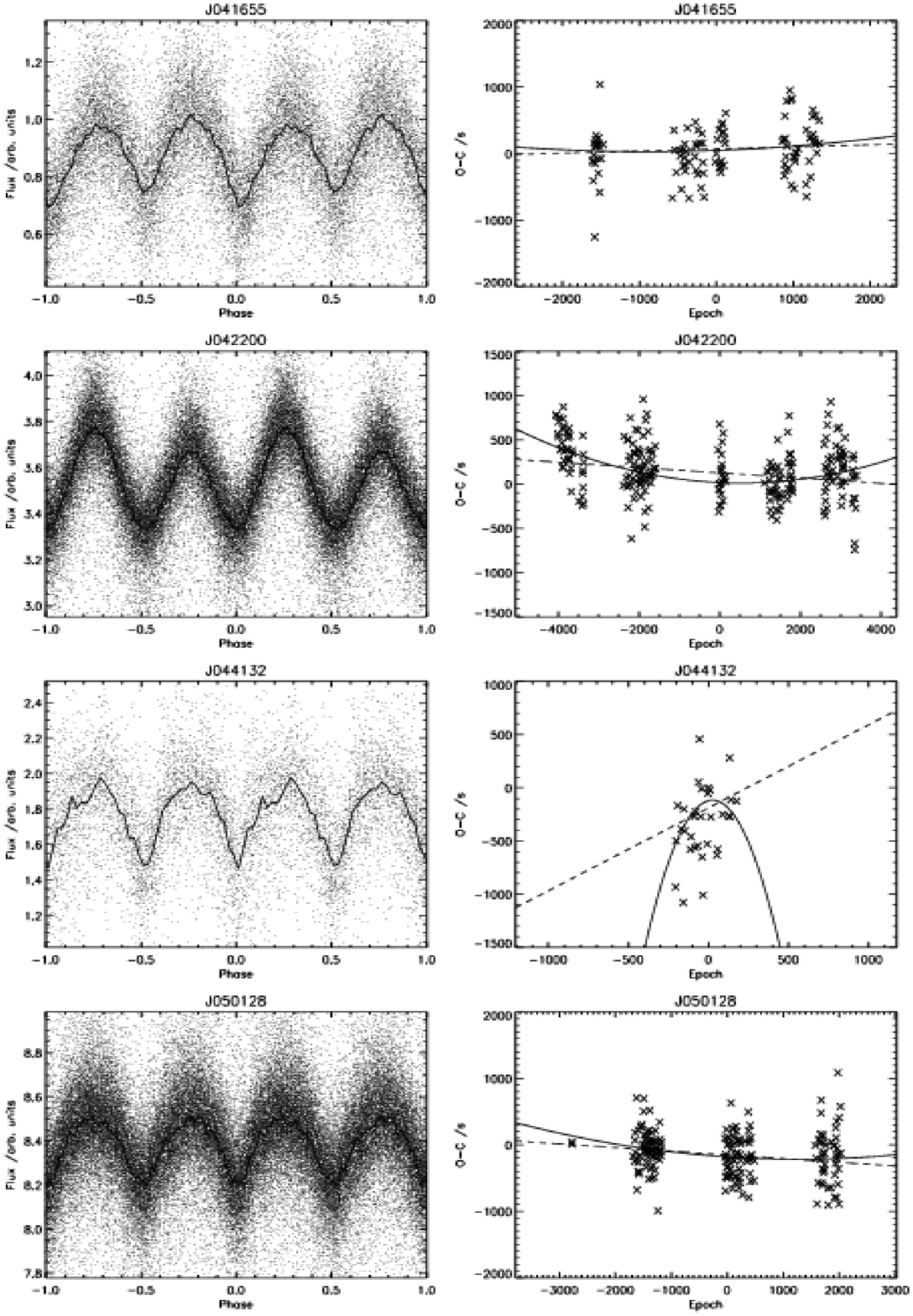}
\end{center}\textbf{Fig.~\ref{appfig1}} continued. \clearpage

\begin{center} \includegraphics[width=16cm]{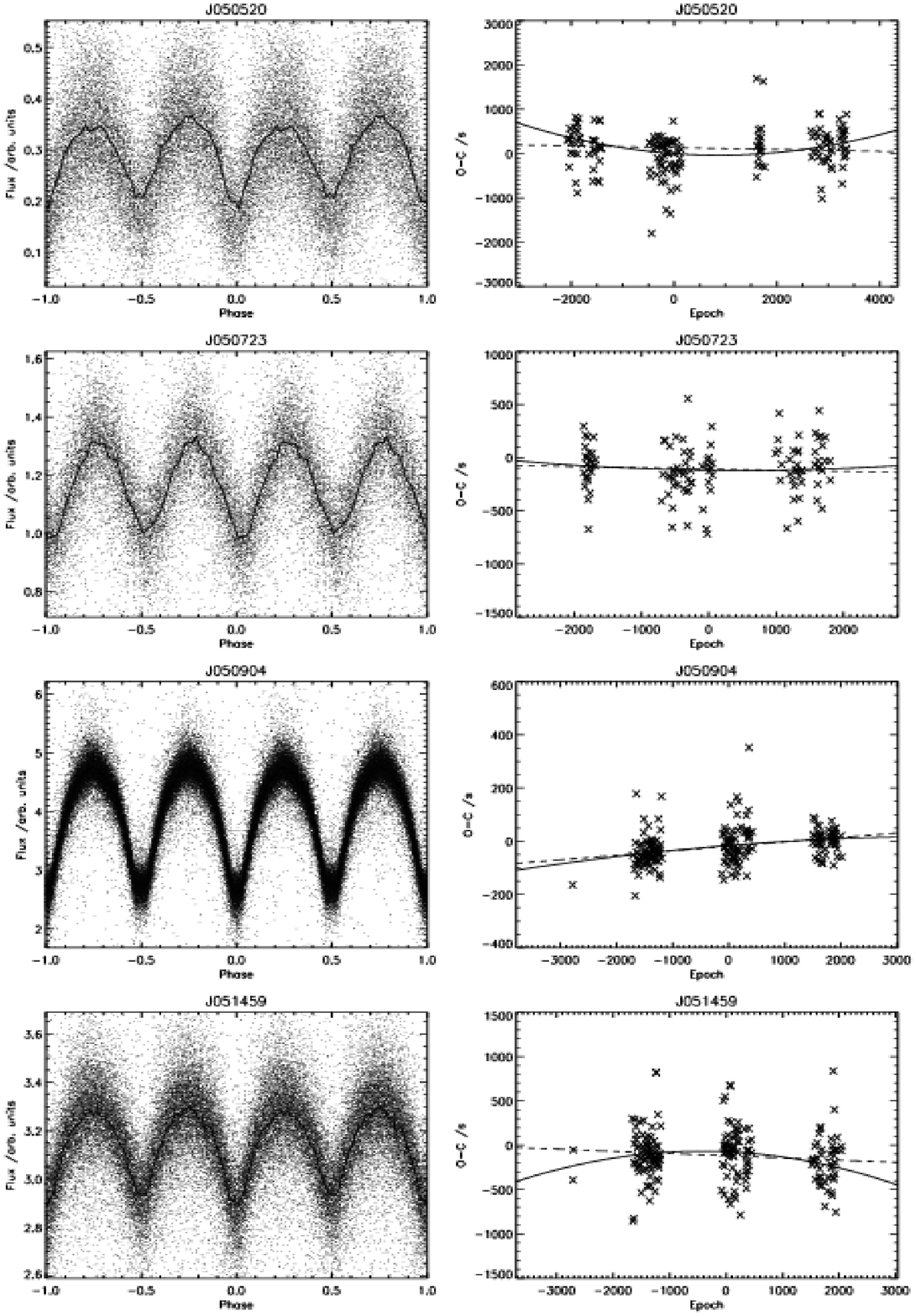}
\end{center}\textbf{Fig.~\ref{appfig1}} continued. \clearpage

\begin{center} \includegraphics[width=16cm]{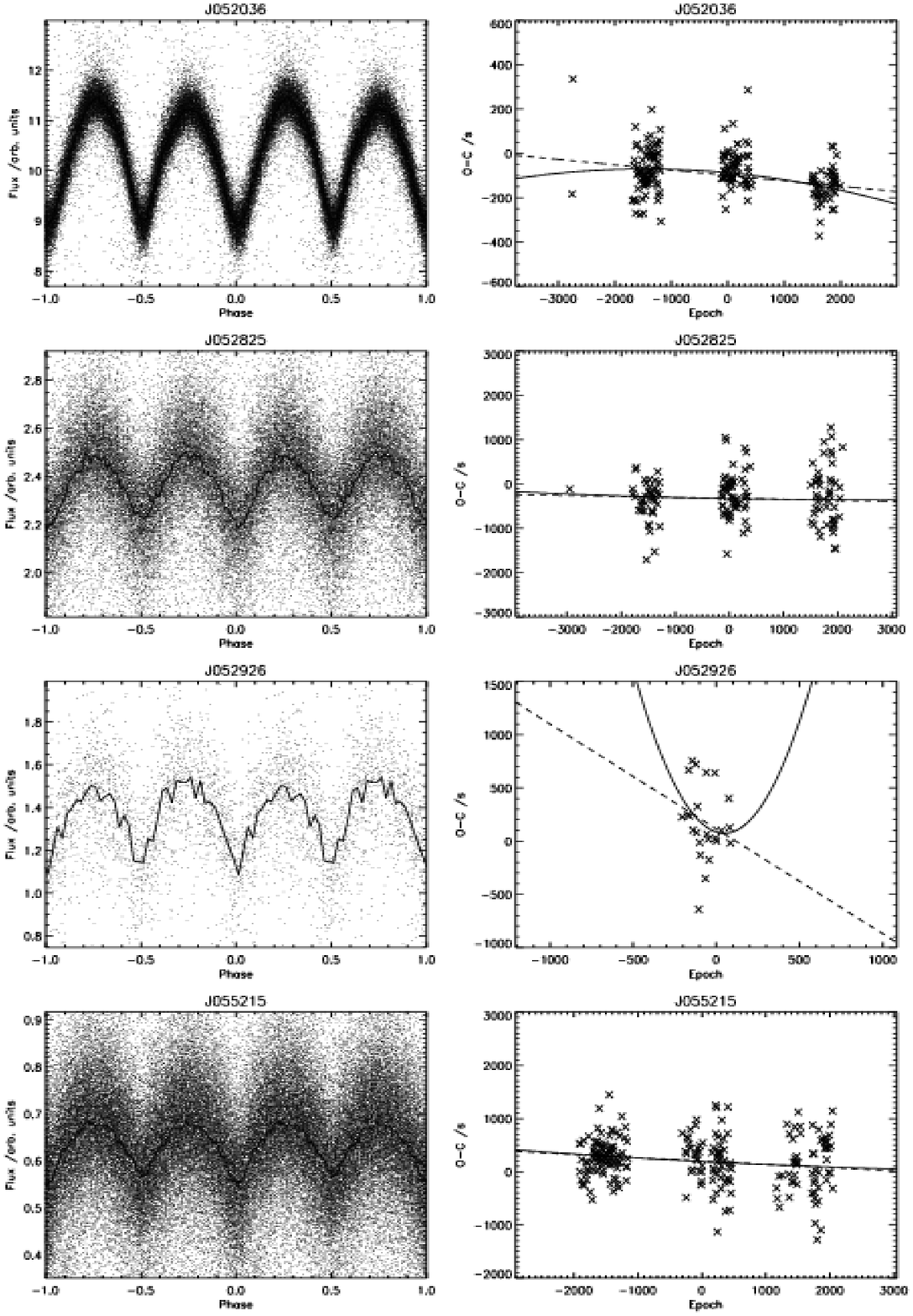}
\end{center}\textbf{Fig.~\ref{appfig1}} continued. \clearpage

\begin{center} \includegraphics[width=16cm]{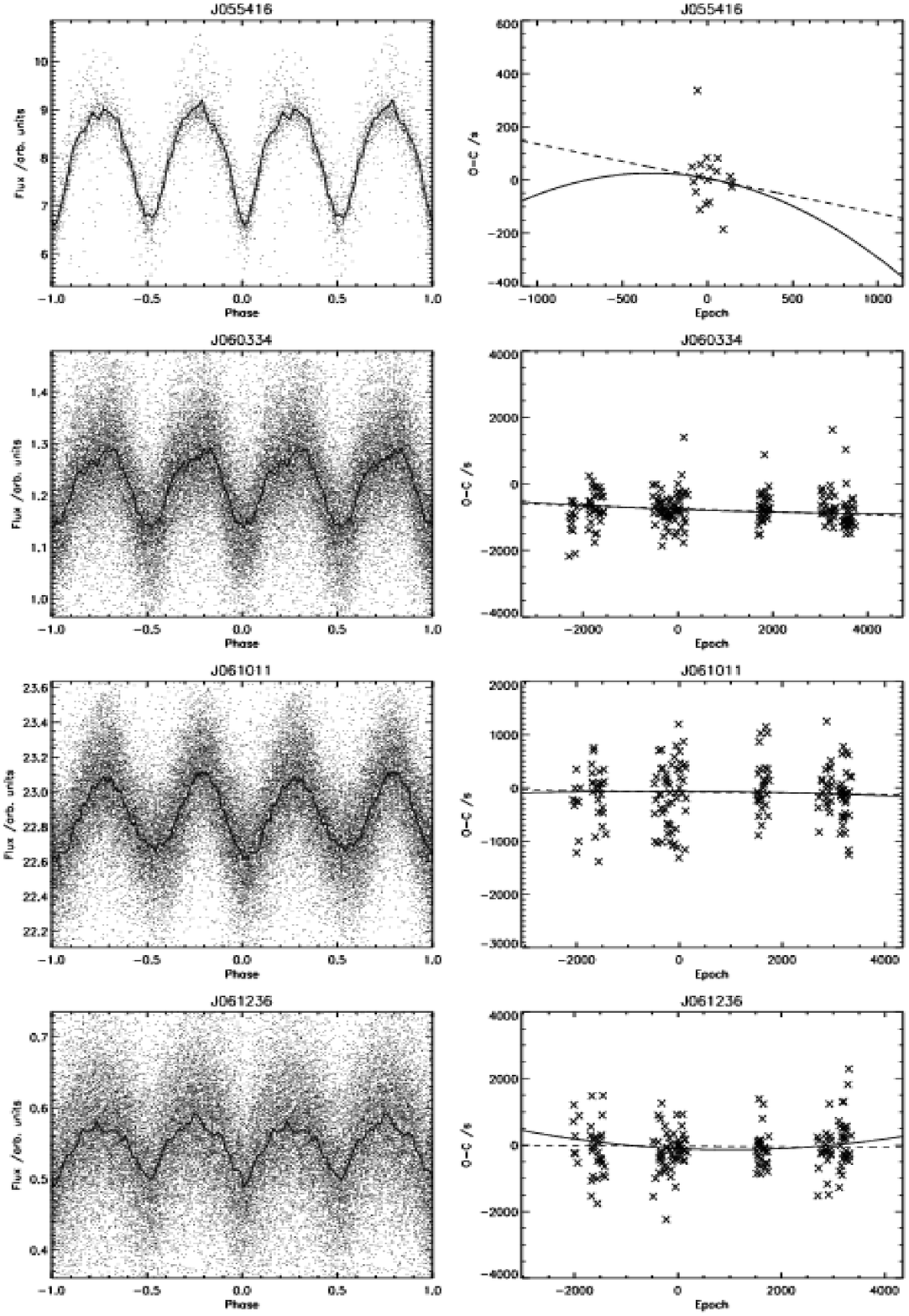}
\end{center}\textbf{Fig.~\ref{appfig1}} continued. \clearpage

\begin{center} \includegraphics[width=16cm]{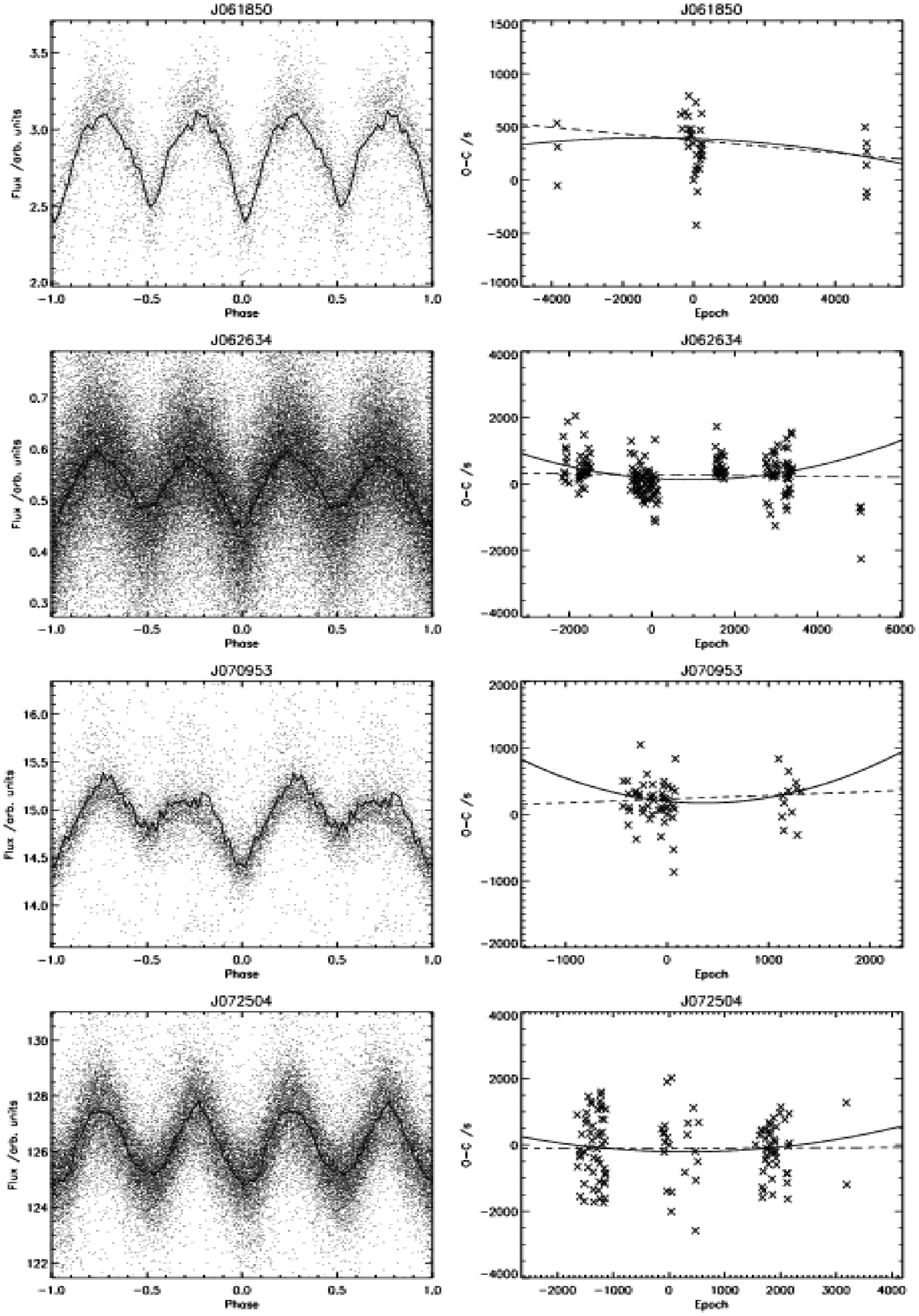}
\end{center}\textbf{Fig.~\ref{appfig1}} continued. \clearpage

\begin{center} \includegraphics[width=16cm]{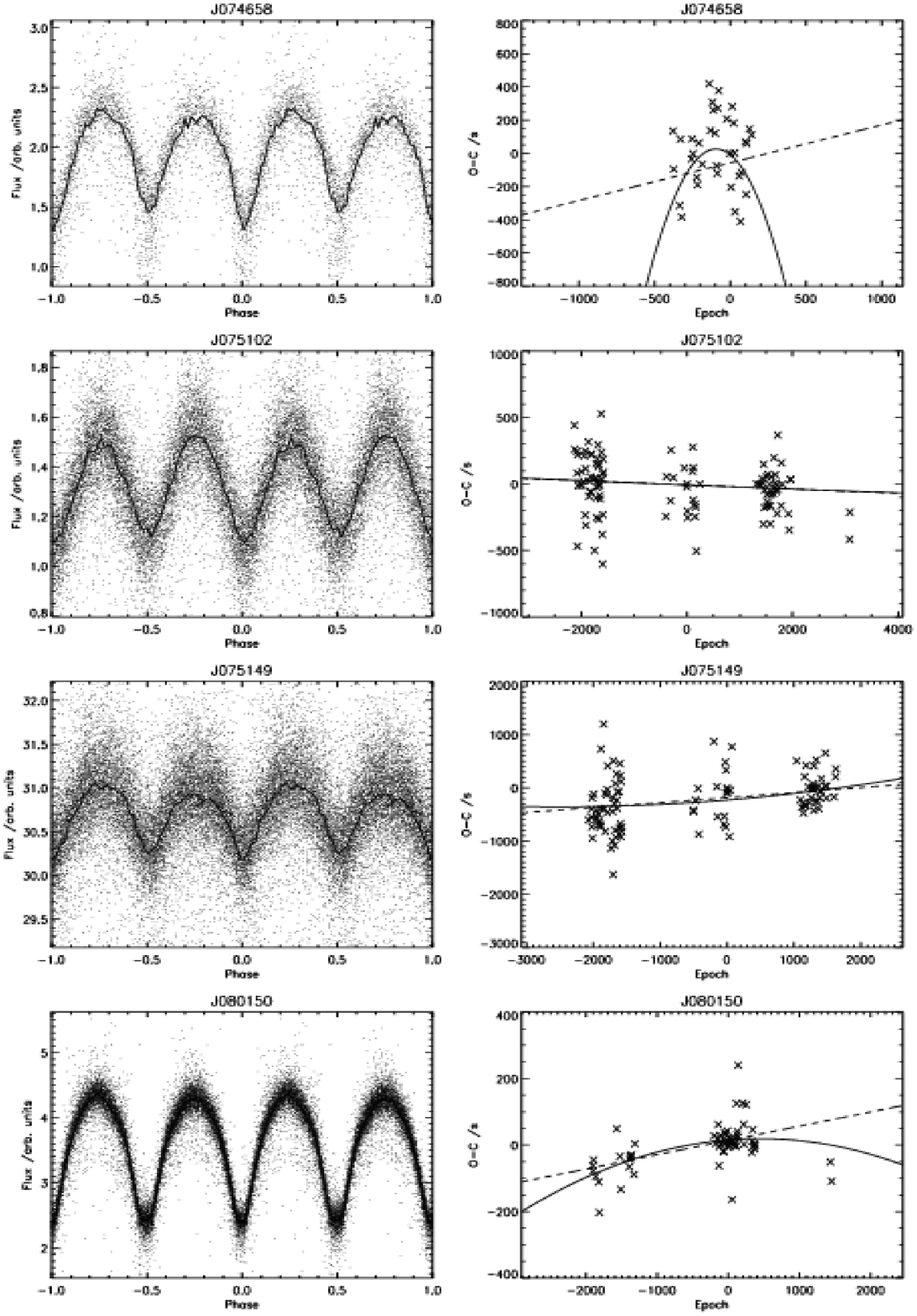}
\end{center}\textbf{Fig.~\ref{appfig1}} continued. \clearpage

\begin{center} \includegraphics[width=16cm]{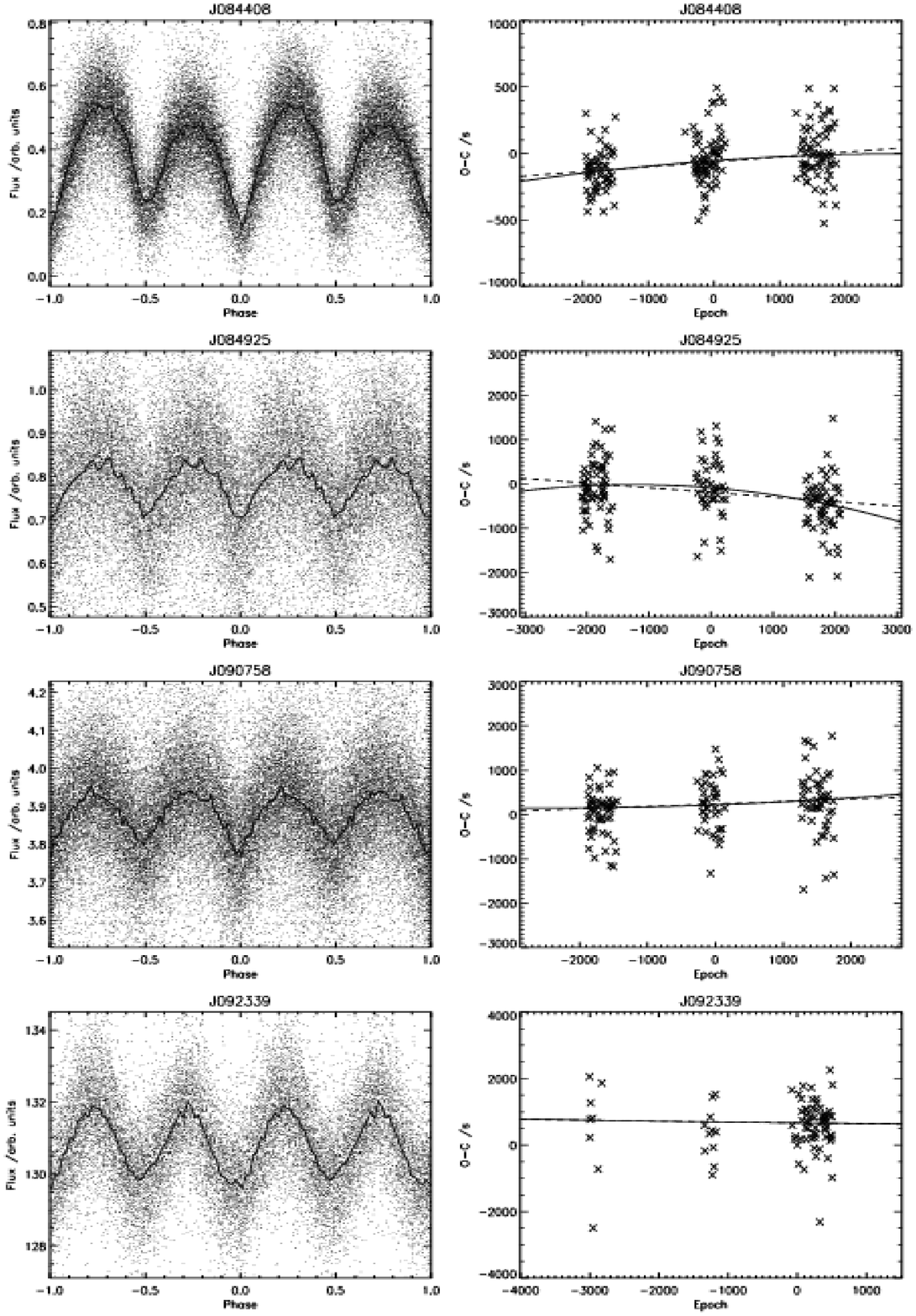}
\end{center}\textbf{Fig.~\ref{appfig1}} continued. \clearpage

\begin{center} \includegraphics[width=16cm]{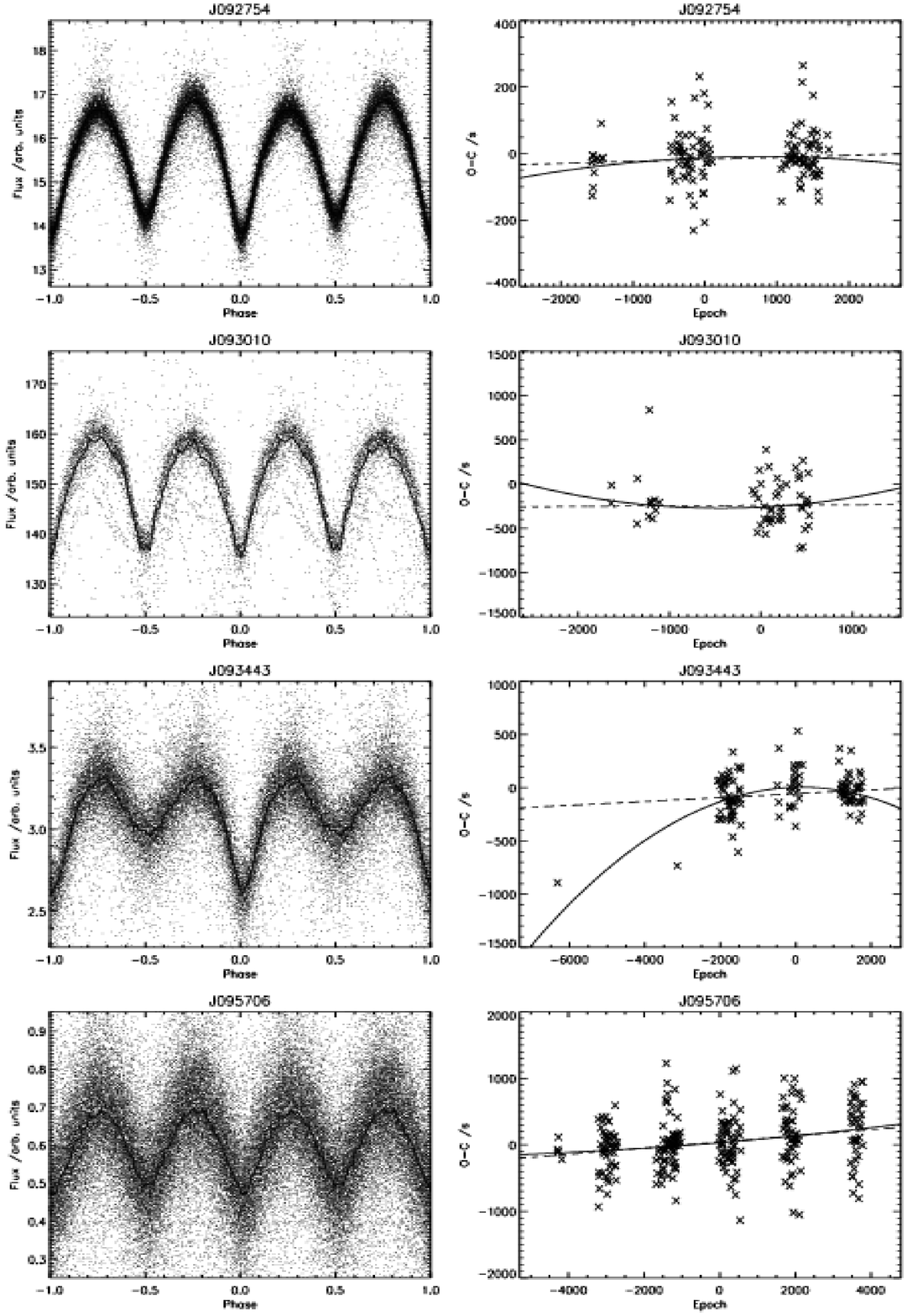}
\end{center}\textbf{Fig.~\ref{appfig1}} continued. \clearpage

\begin{center} \includegraphics[width=16cm]{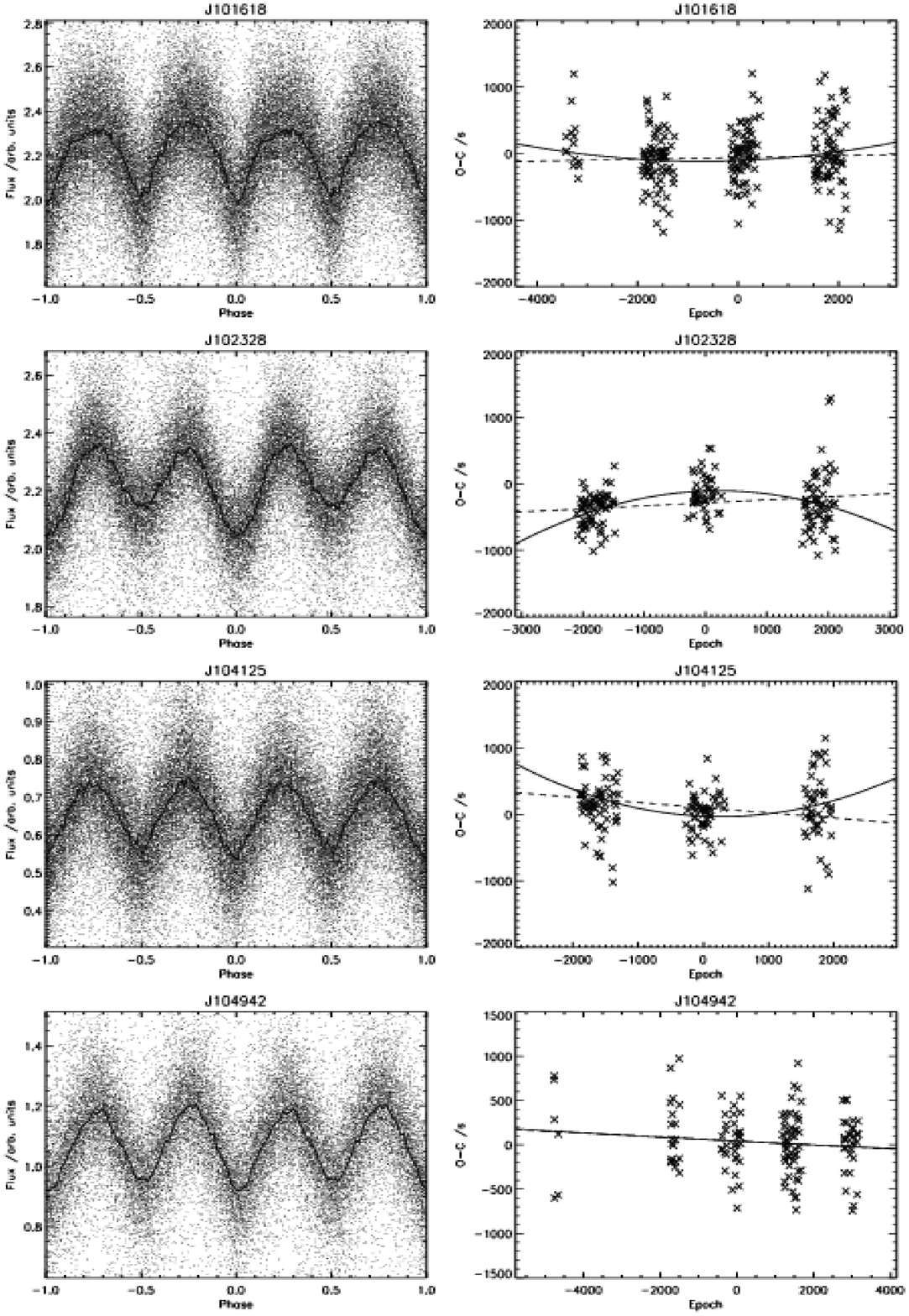}
\end{center}\textbf{Fig.~\ref{appfig1}} continued. \clearpage

\begin{center} \includegraphics[width=16cm]{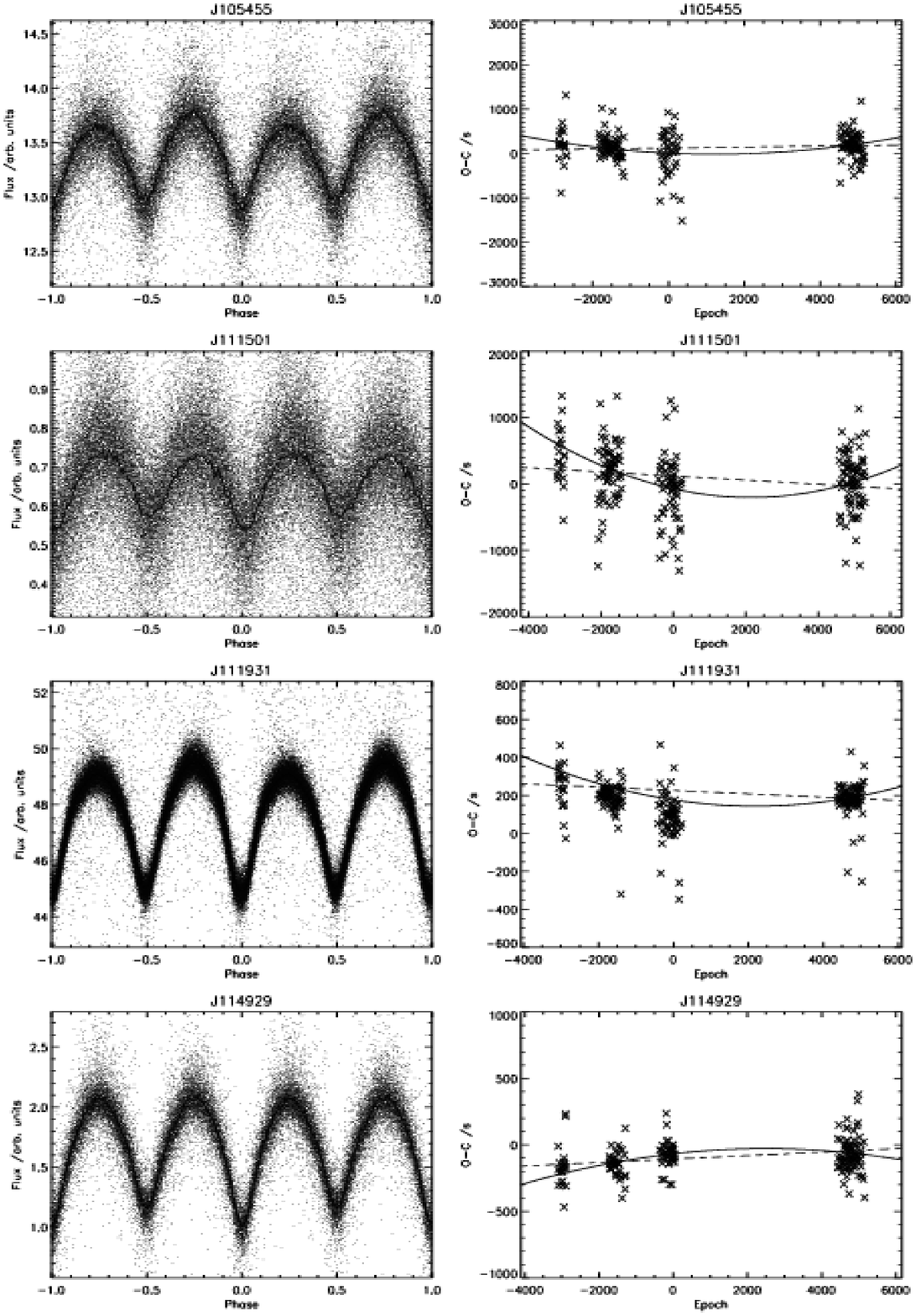}
\end{center}\textbf{Fig.~\ref{appfig1}} continued. \clearpage

\begin{center} \includegraphics[width=16cm]{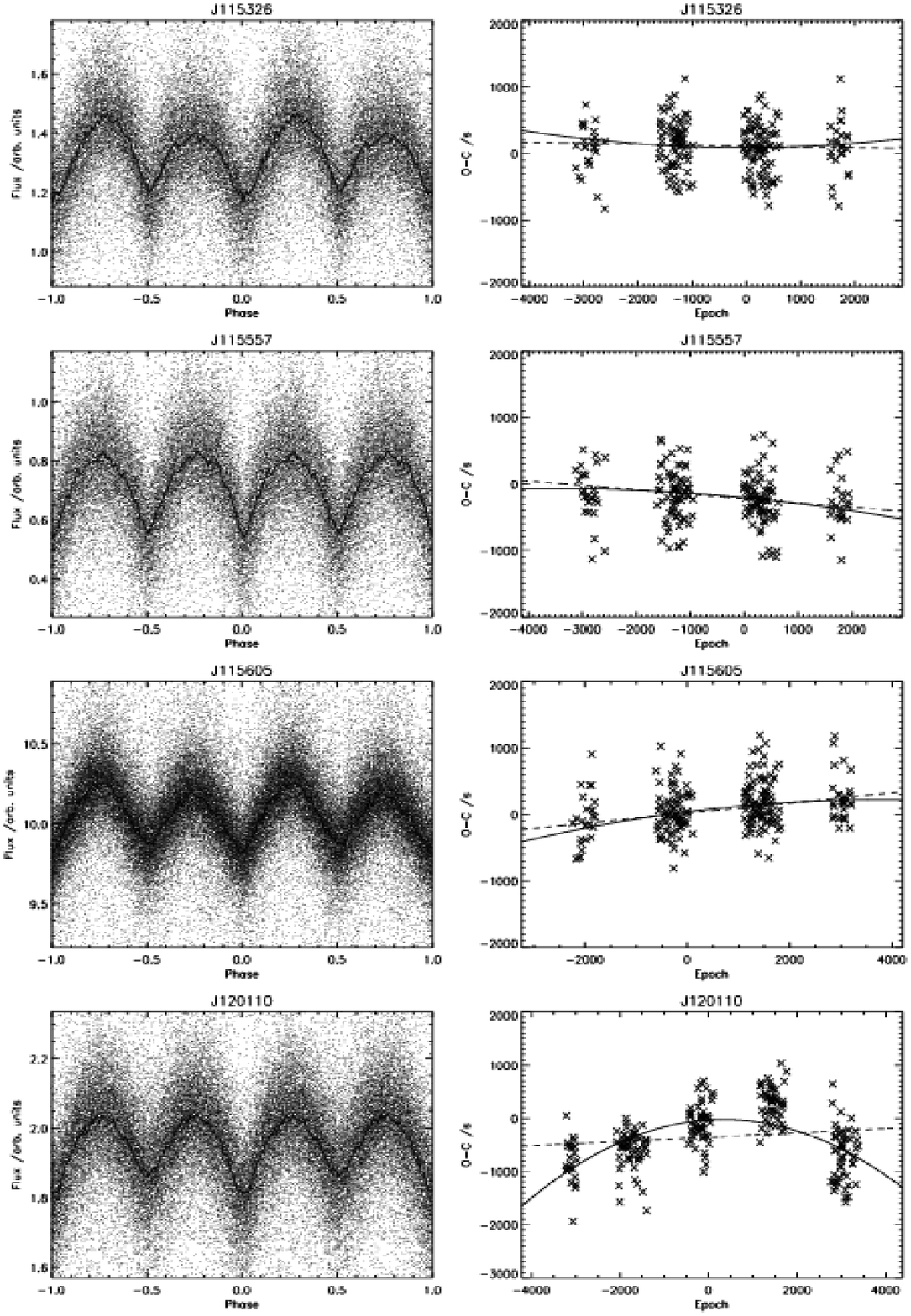}
\end{center}\textbf{Fig.~\ref{appfig1}} continued. \clearpage

\begin{center} \includegraphics[width=16cm]{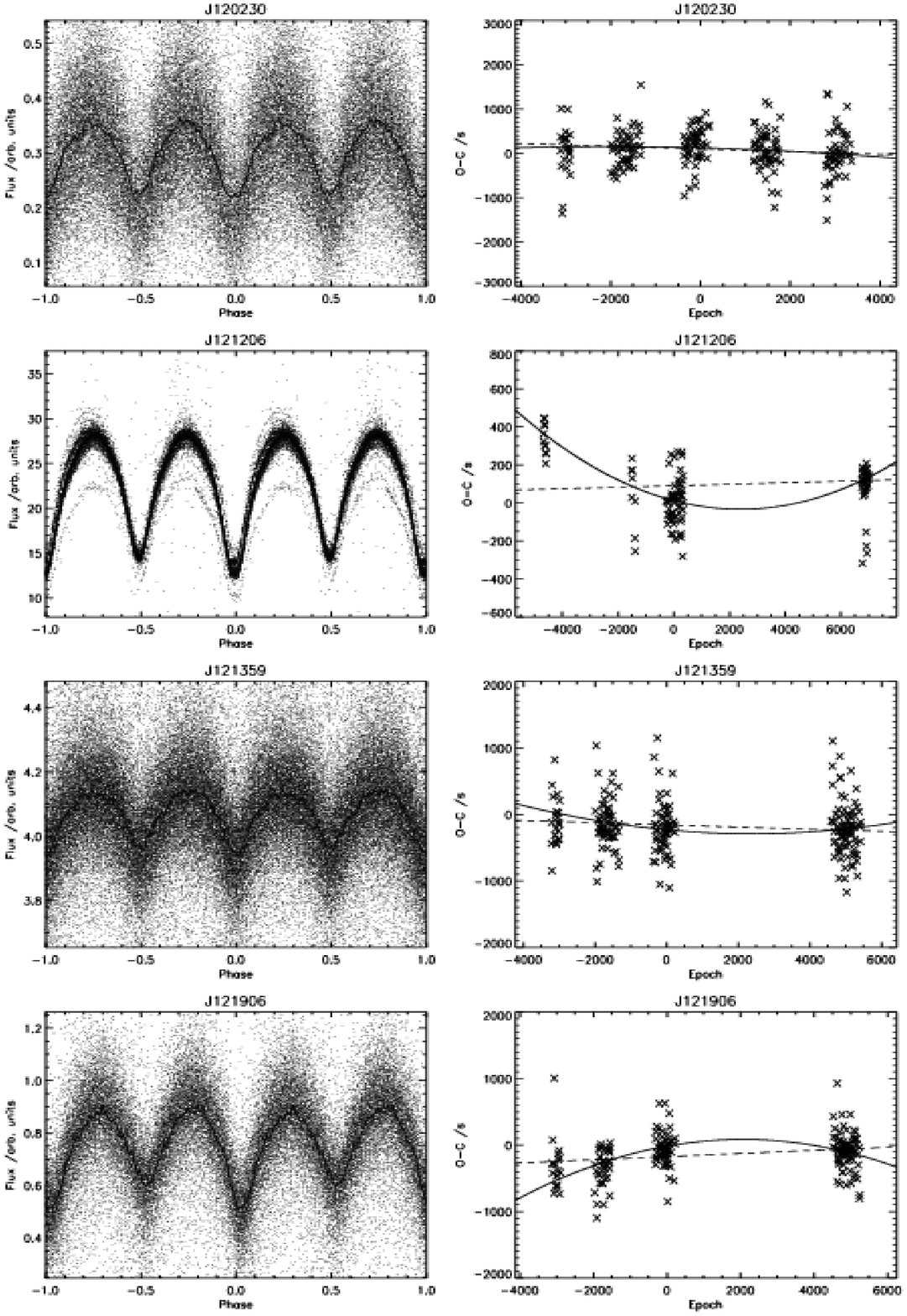}
\end{center}\textbf{Fig.~\ref{appfig1}} continued. \clearpage

\begin{center} \includegraphics[width=16cm]{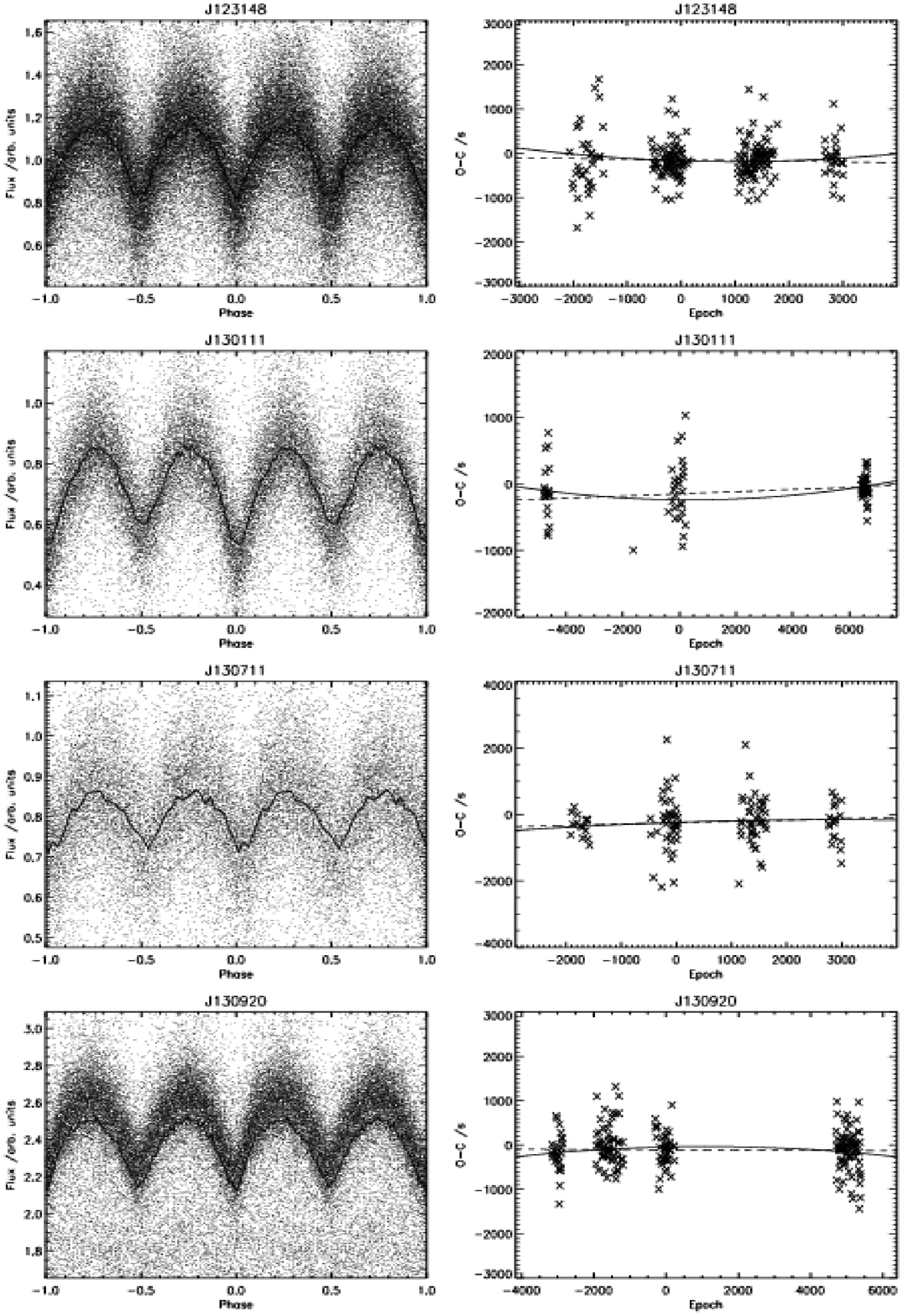}
\end{center}\textbf{Fig.~\ref{appfig1}} continued. \clearpage

\begin{center} \includegraphics[width=16cm]{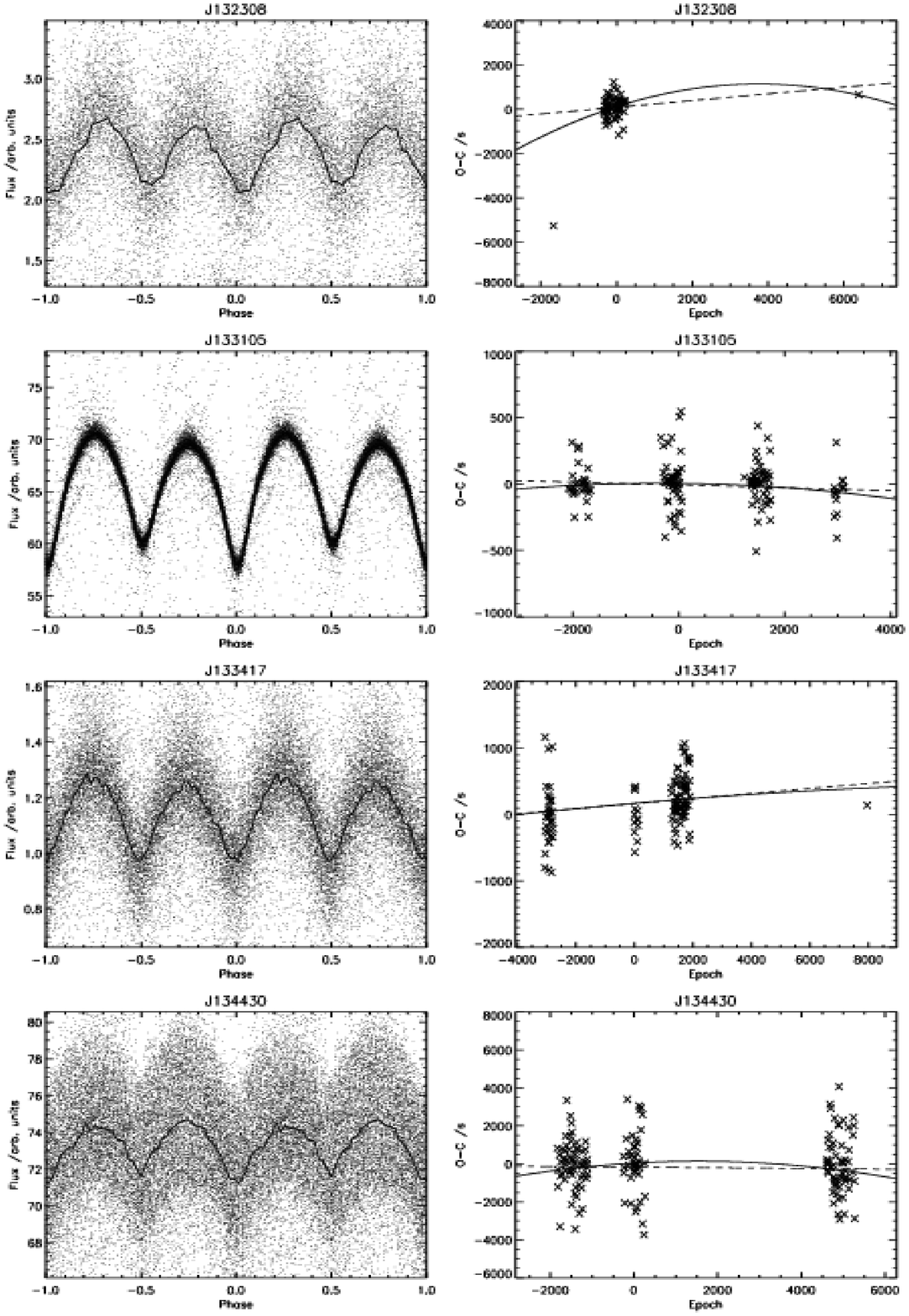}
\end{center}\textbf{Fig.~\ref{appfig1}} continued. \clearpage

\begin{center} \includegraphics[width=16cm]{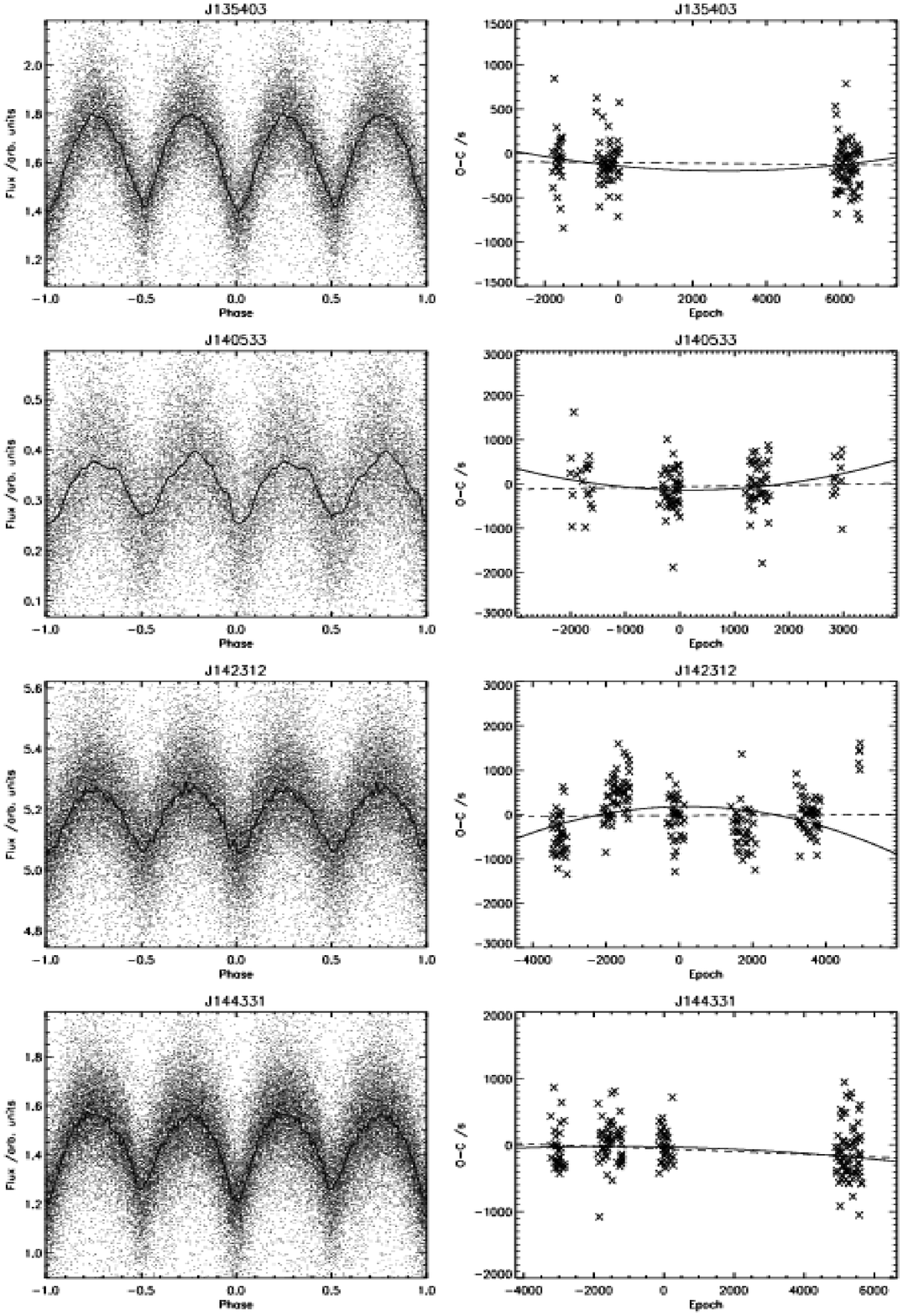}
\end{center}\textbf{Fig.~\ref{appfig1}} continued. \clearpage

\begin{center} \includegraphics[width=16cm]{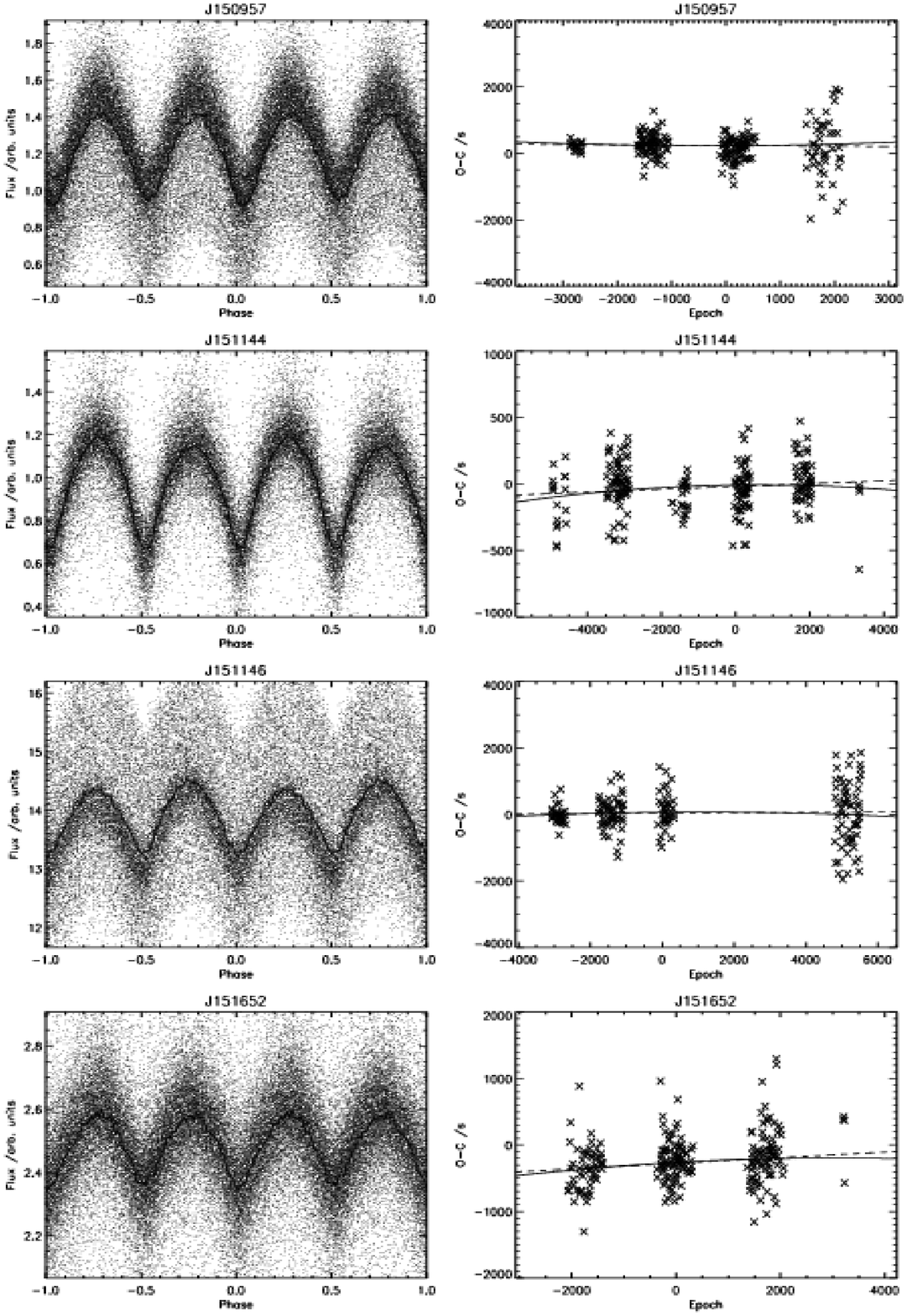}
\end{center}\textbf{Fig.~\ref{appfig1}} continued. \clearpage

\begin{center} \includegraphics[width=16cm]{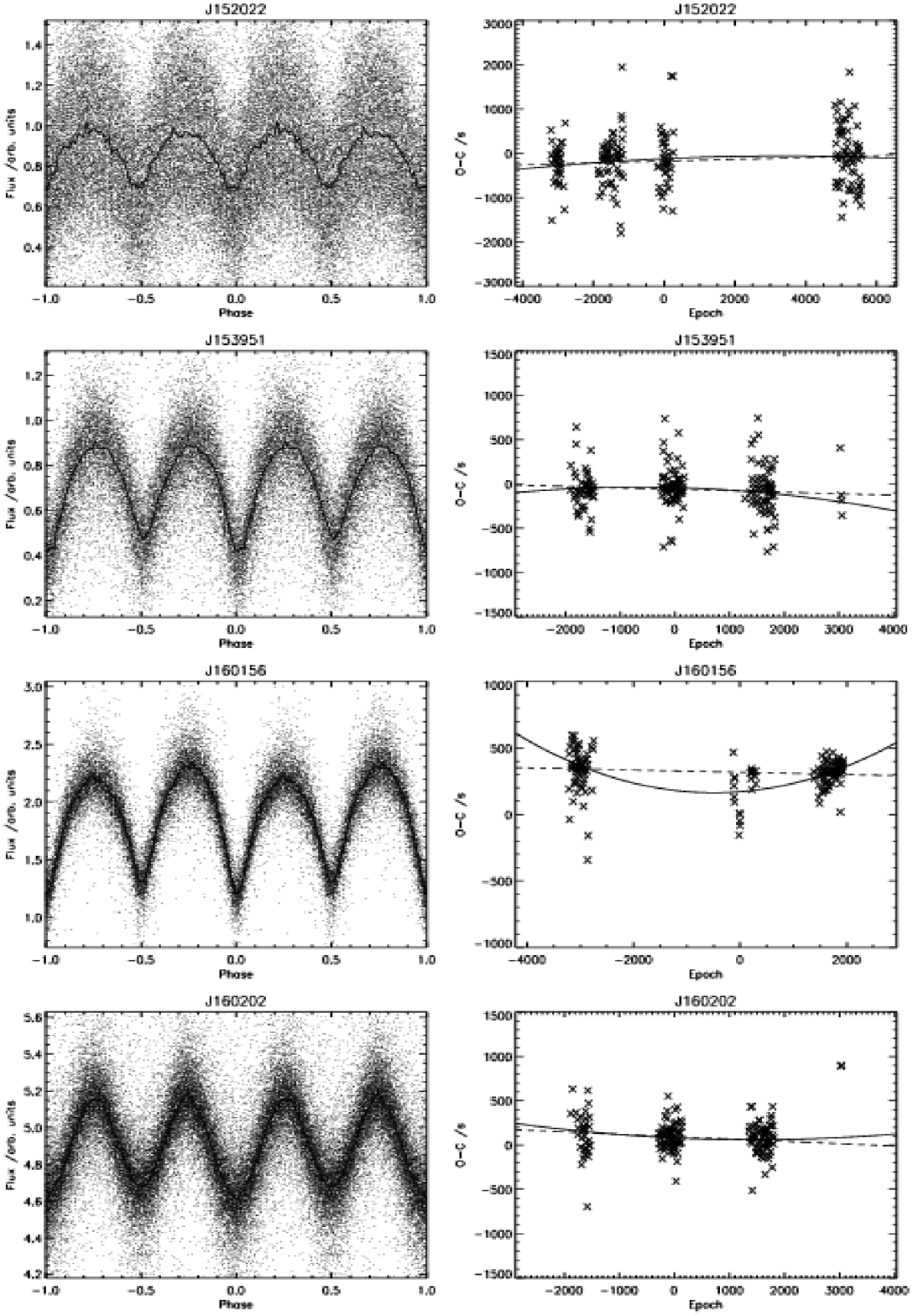}
\end{center}\textbf{Fig.~\ref{appfig1}} continued. \clearpage

\begin{center} \includegraphics[width=16cm]{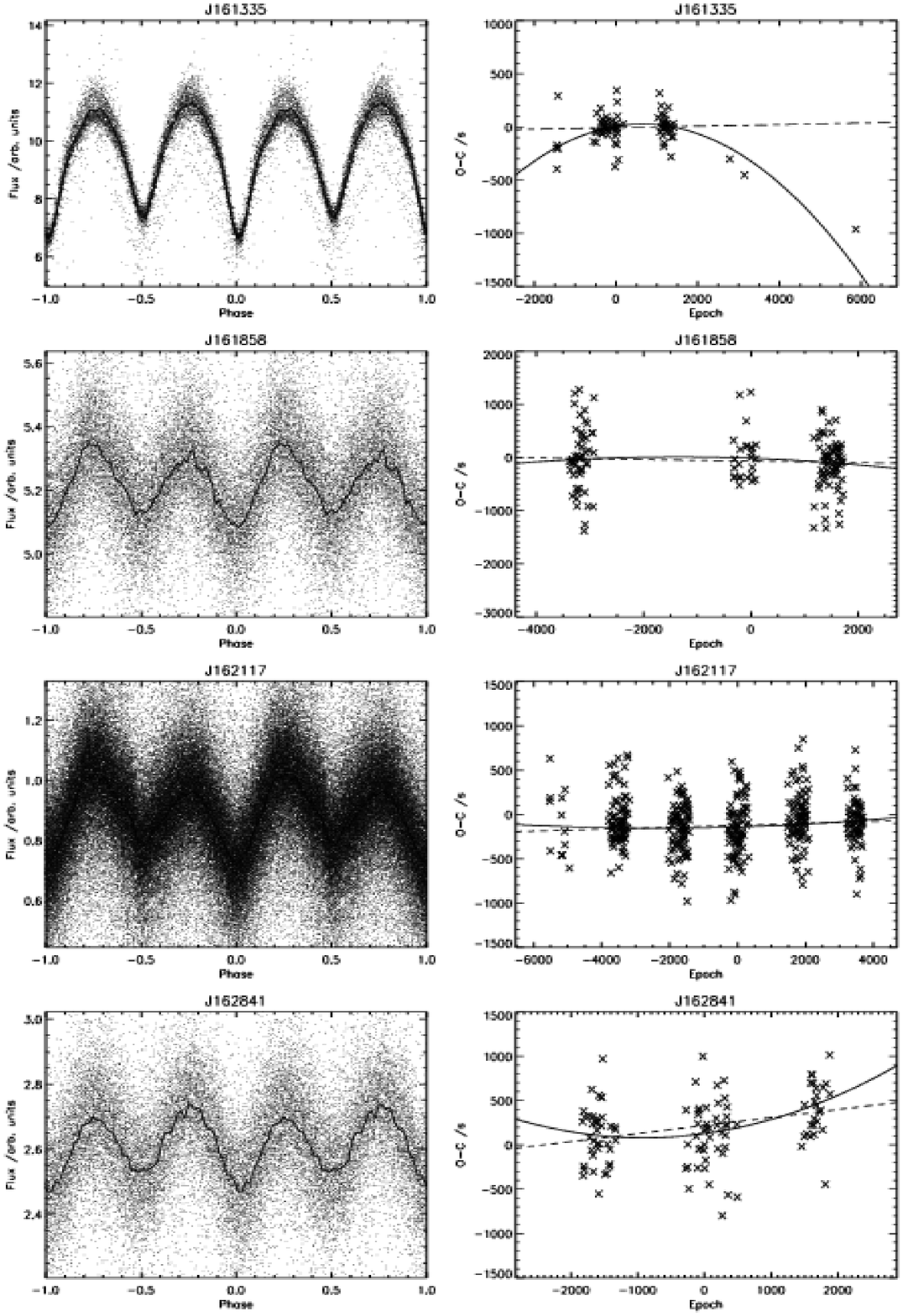}
\end{center}\textbf{Fig.~\ref{appfig1}} continued. \clearpage

\begin{center} \includegraphics[width=16cm]{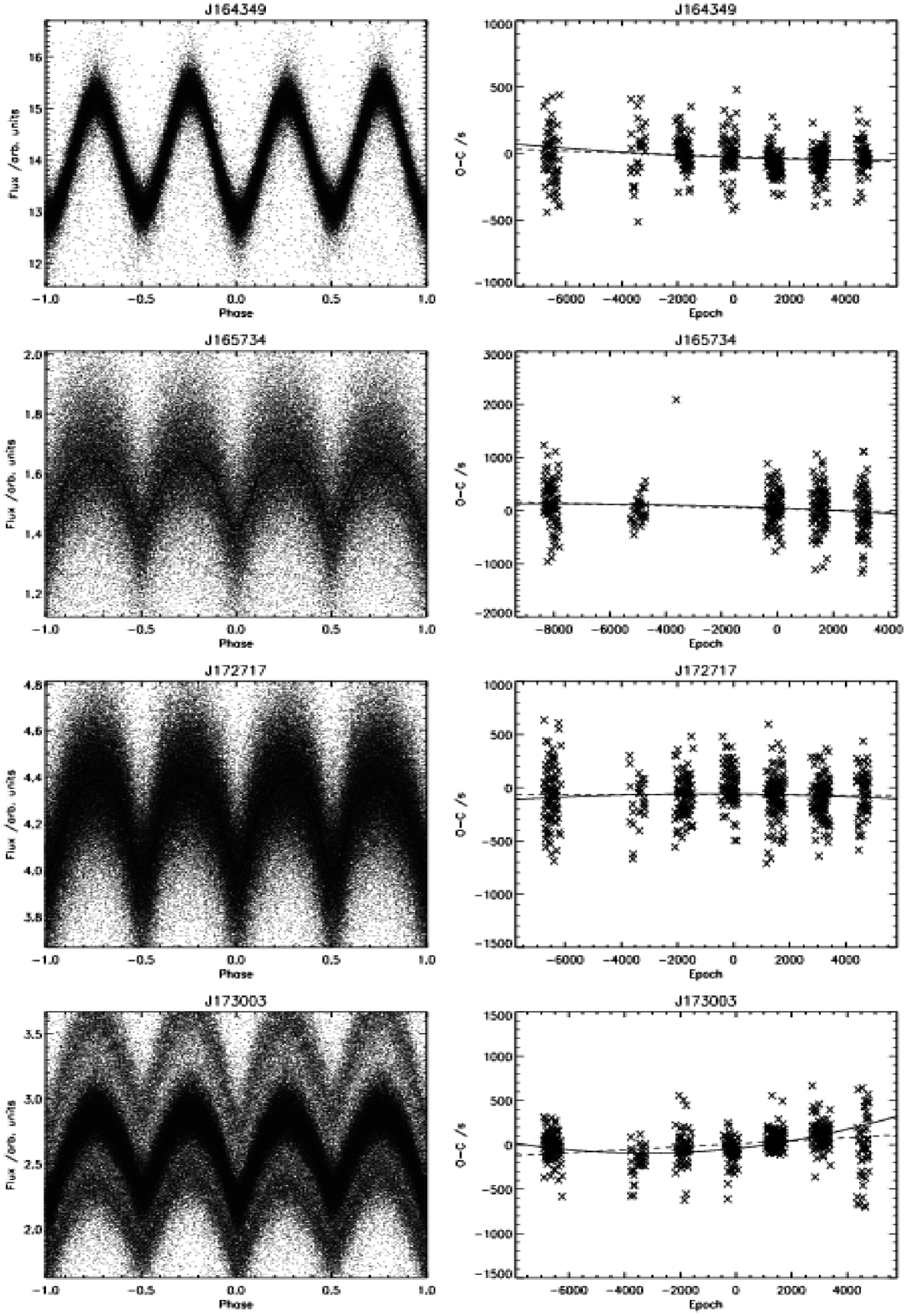}
\end{center}\textbf{Fig.~\ref{appfig1}} continued. \clearpage

\begin{center} \includegraphics[width=16cm]{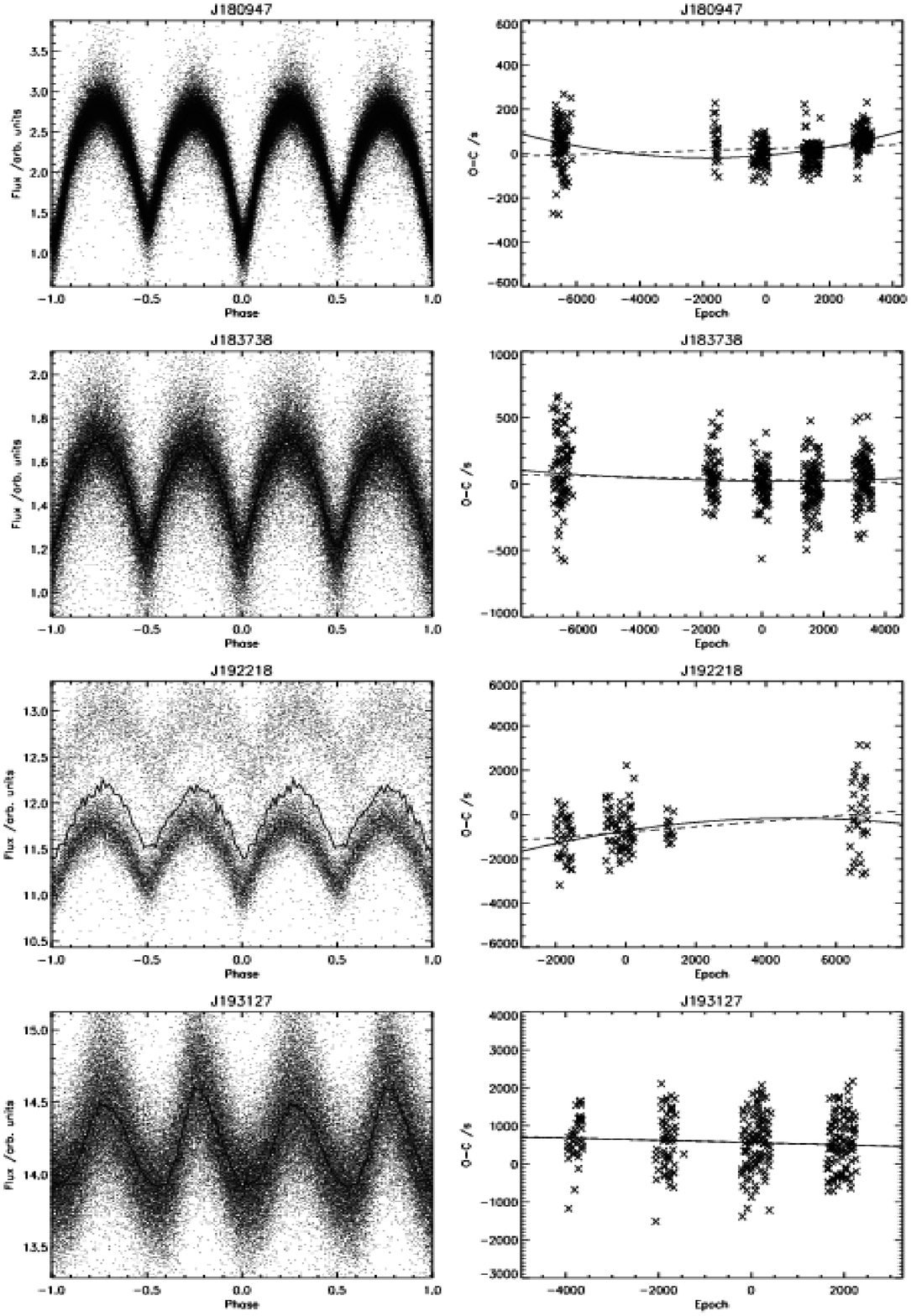}
\end{center}\textbf{Fig.~\ref{appfig1}} continued. \clearpage

\begin{center} \includegraphics[width=16cm]{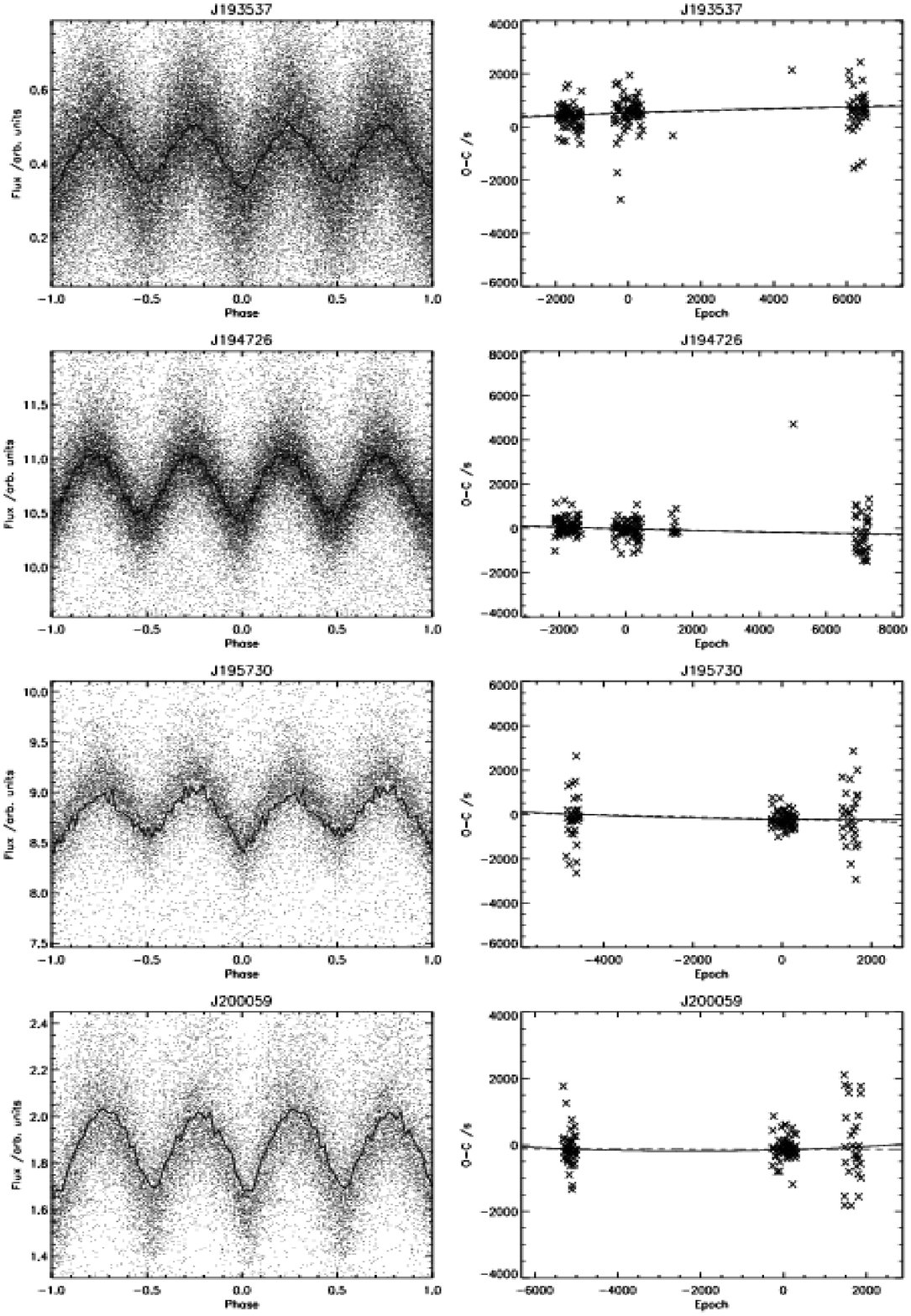}
\end{center}\textbf{Fig.~\ref{appfig1}} continued. \clearpage

\begin{center} \includegraphics[width=16cm]{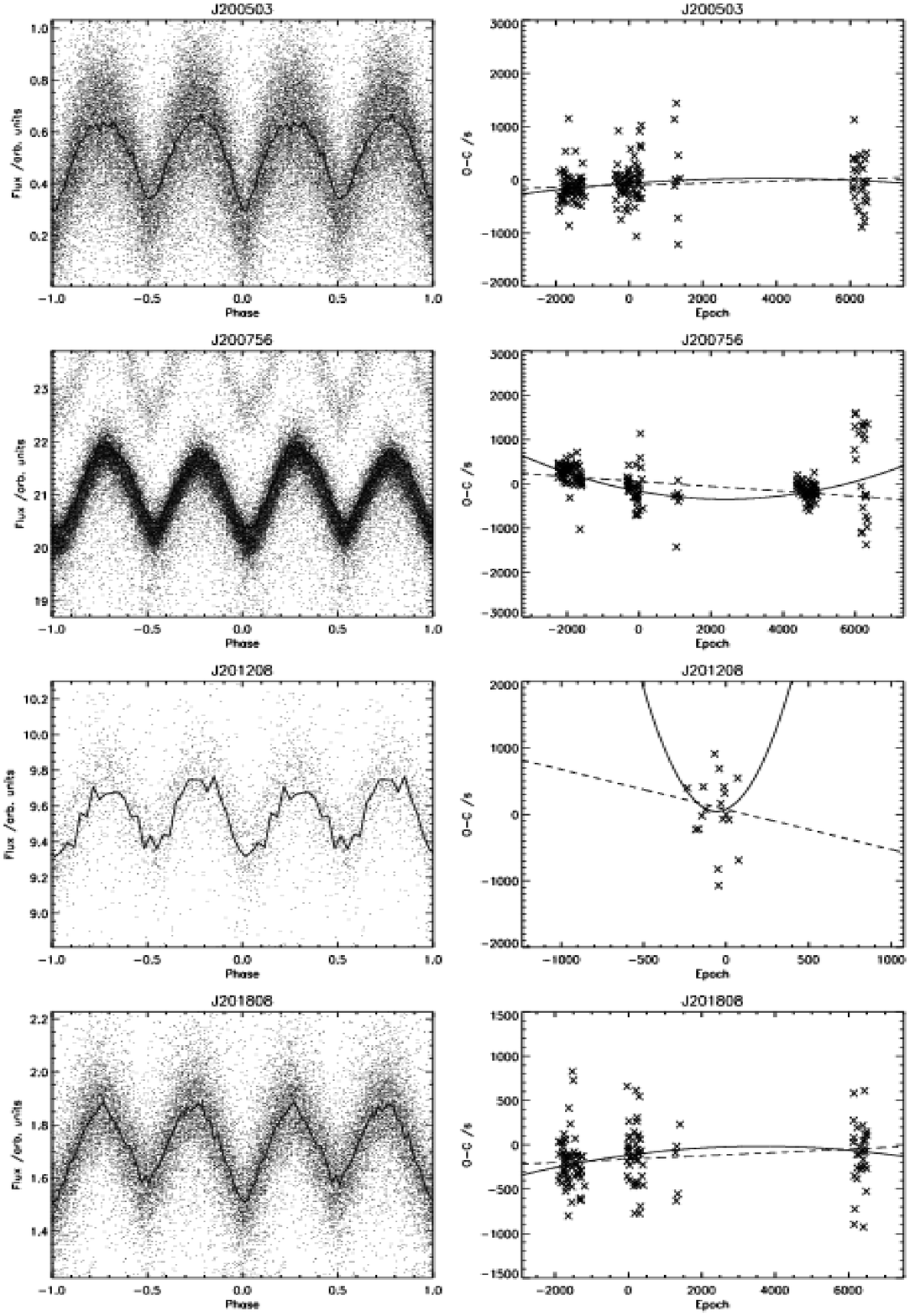}
\end{center}\textbf{Fig.~\ref{appfig1}} continued. \clearpage

\begin{center} \includegraphics[width=16cm]{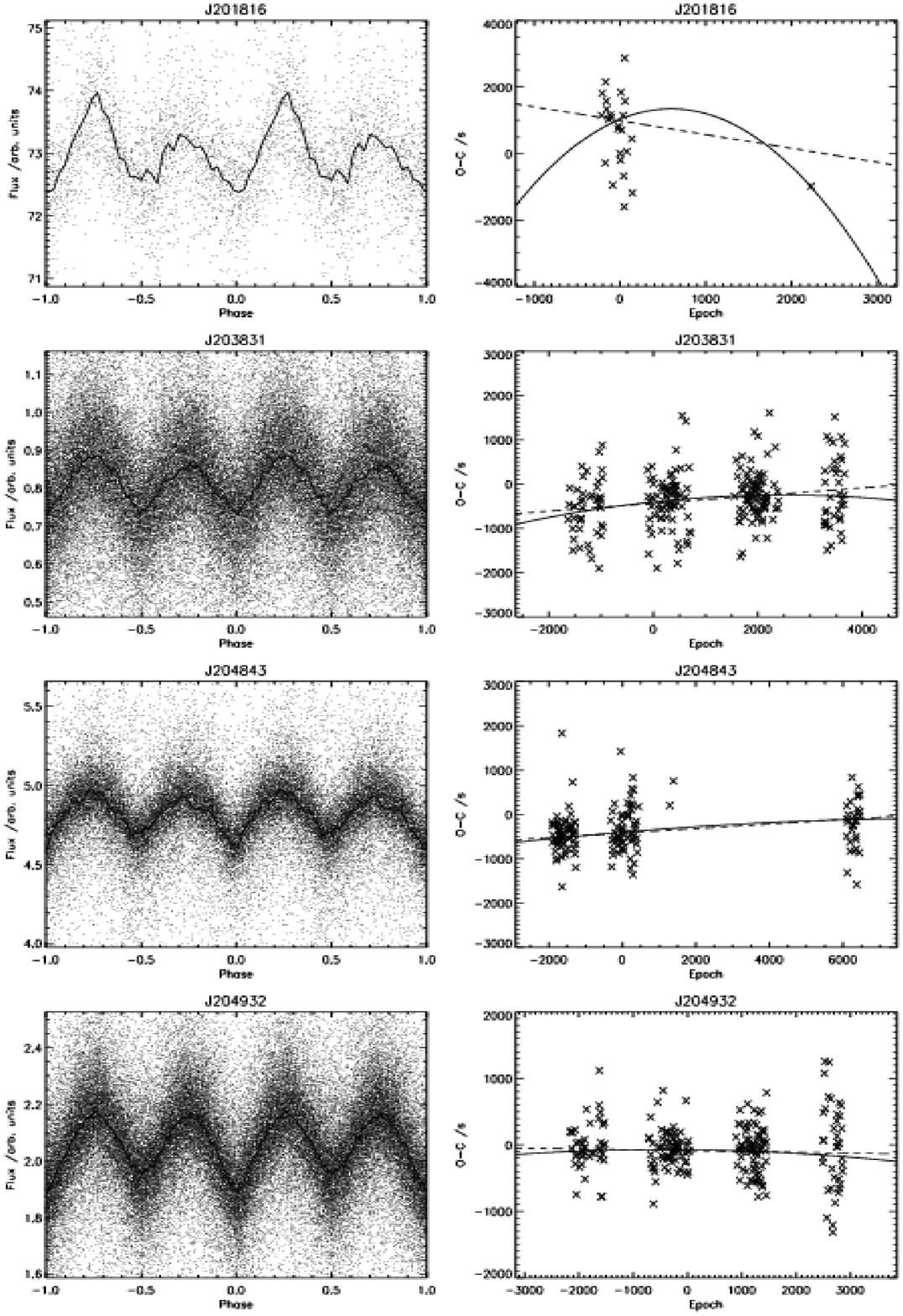}
\end{center}\textbf{Fig.~\ref{appfig1}} continued. \clearpage

\begin{center} \includegraphics[width=16cm]{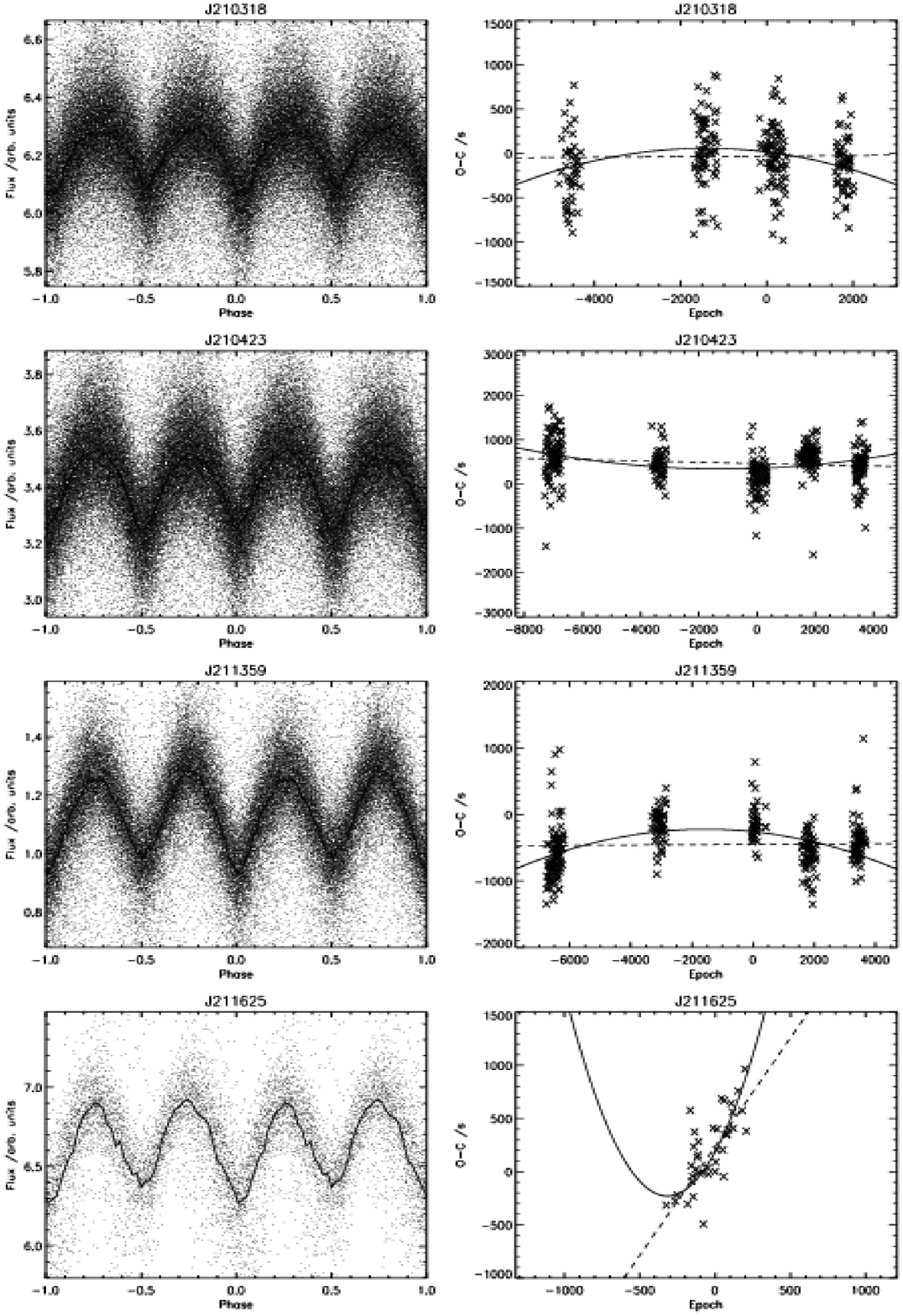}
\end{center}\textbf{Fig.~\ref{appfig1}} continued. \clearpage

\begin{center} \includegraphics[width=16cm]{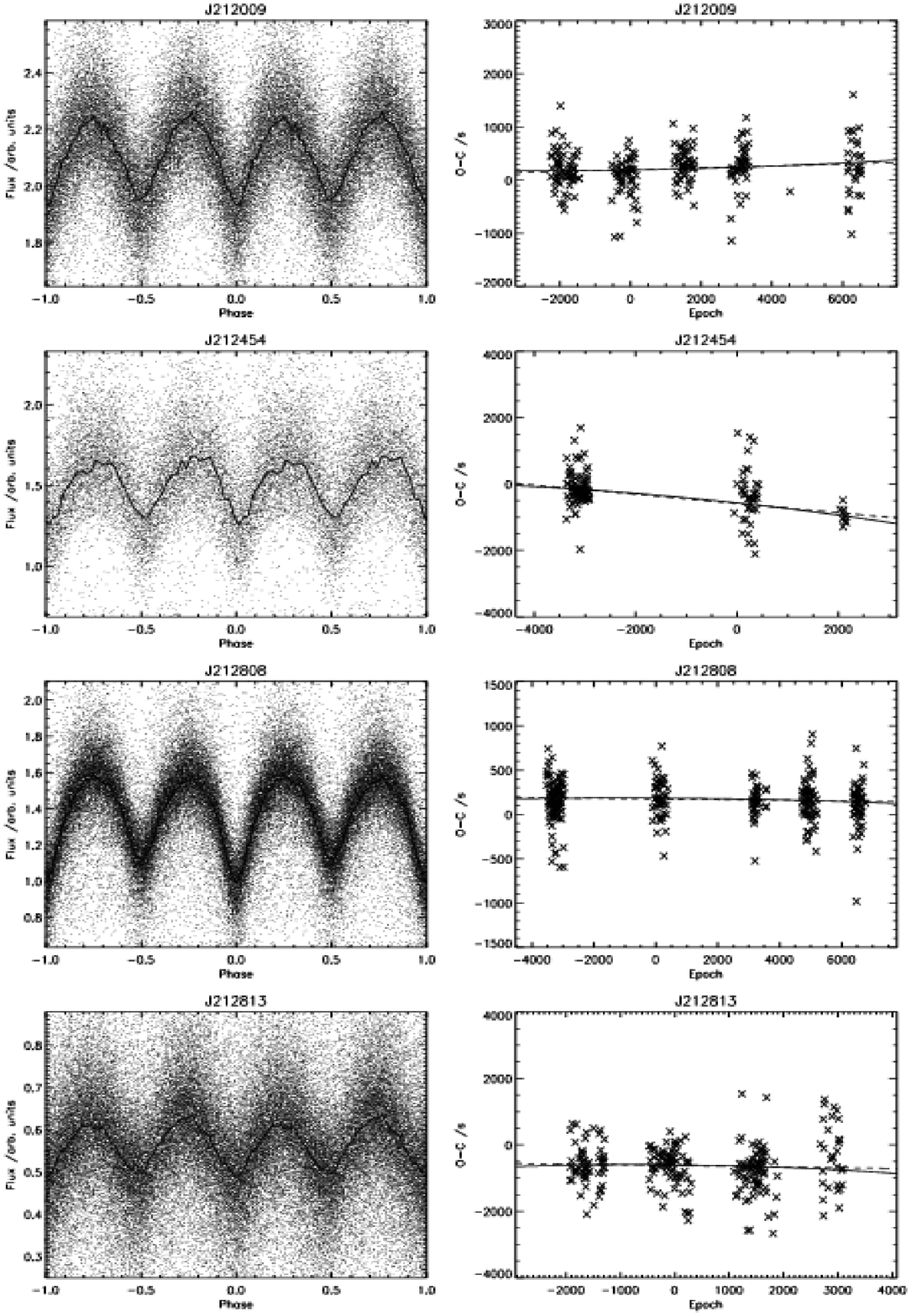}
\end{center}\textbf{Fig.~\ref{appfig1}} continued. \clearpage

\begin{center} \includegraphics[width=16cm]{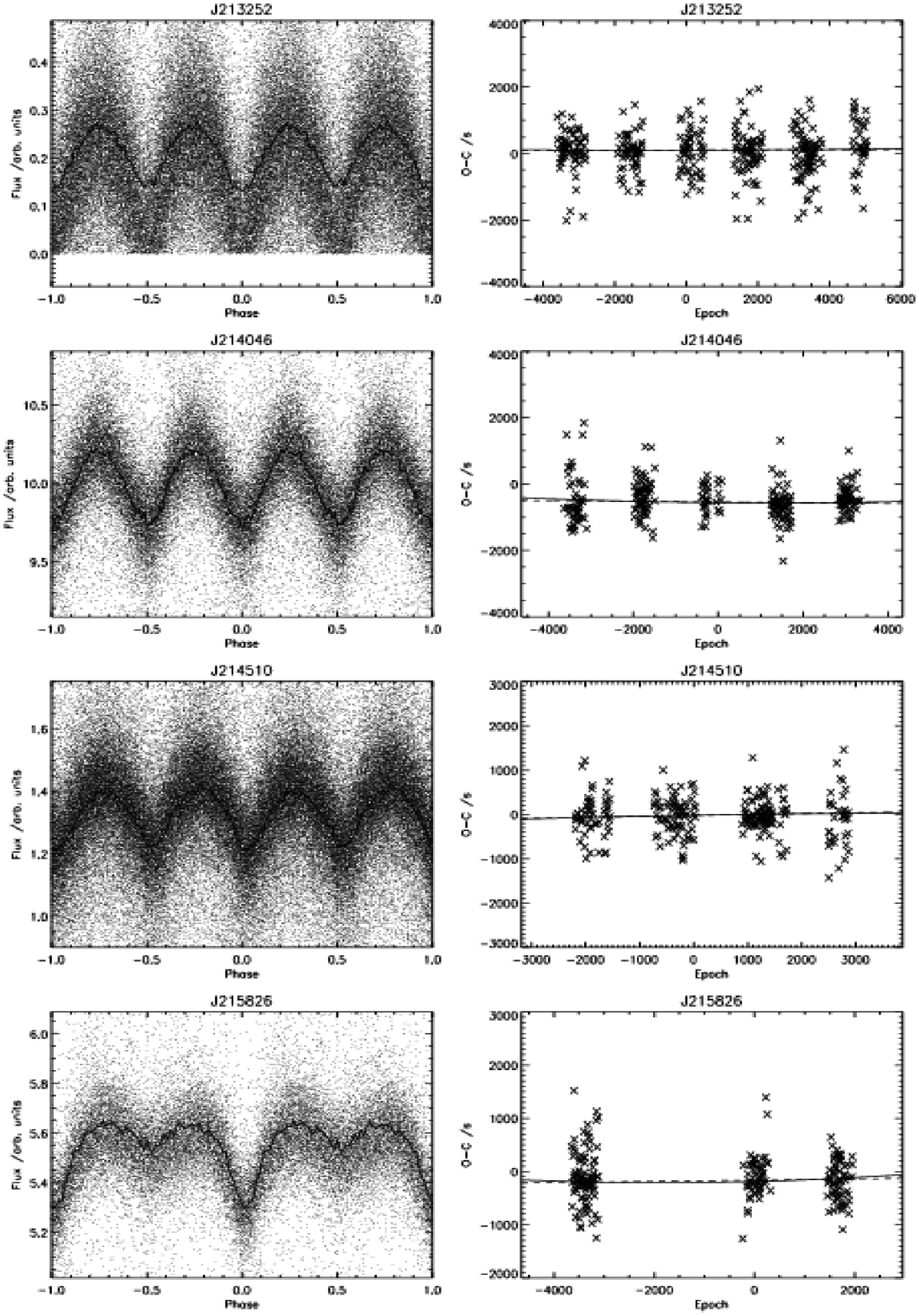}
\end{center}\textbf{Fig.~\ref{appfig1}} continued. \clearpage

\begin{center} \includegraphics[width=16cm]{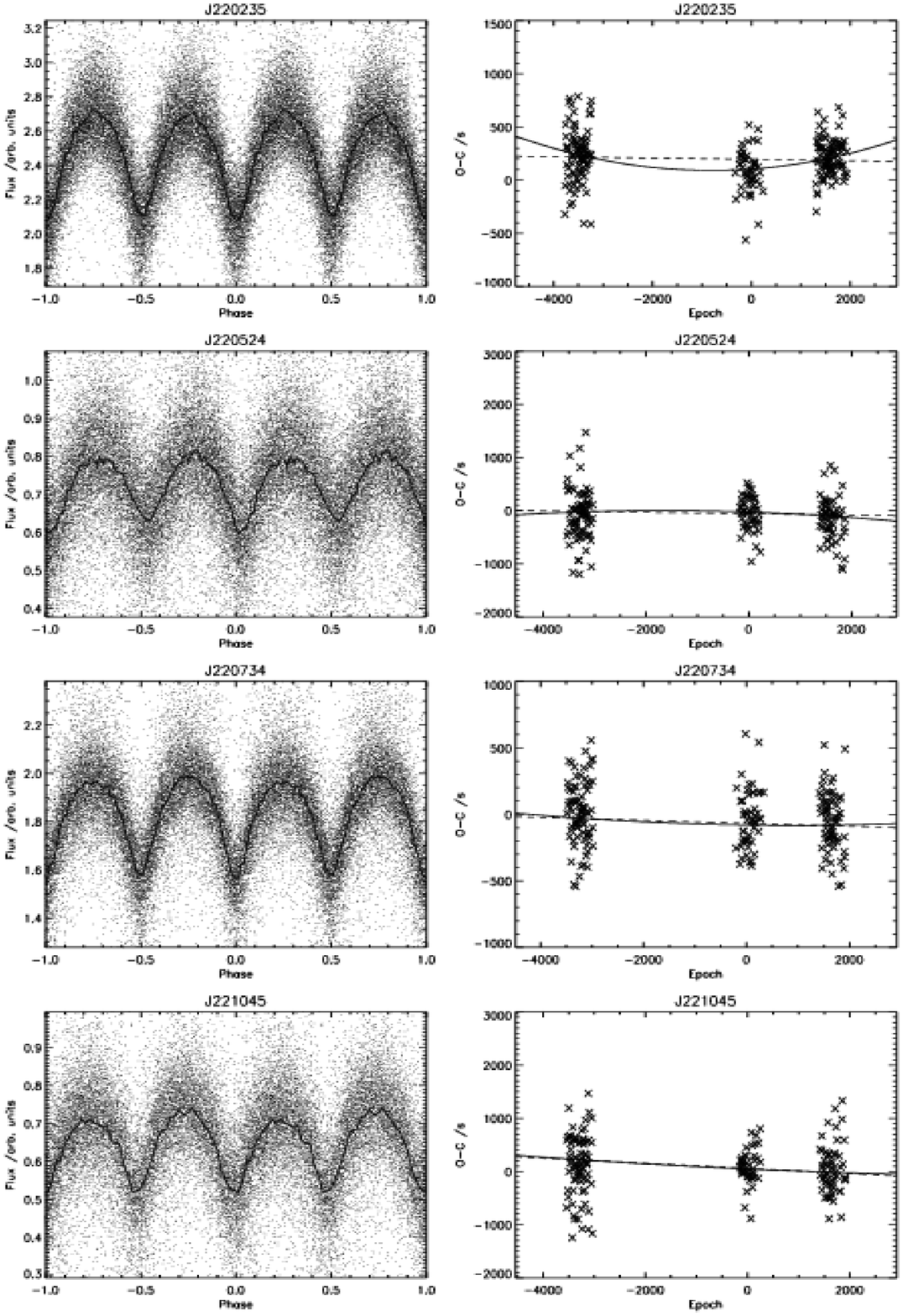}
\end{center}\textbf{Fig.~\ref{appfig1}} continued. \clearpage

\begin{center} \includegraphics[width=16cm]{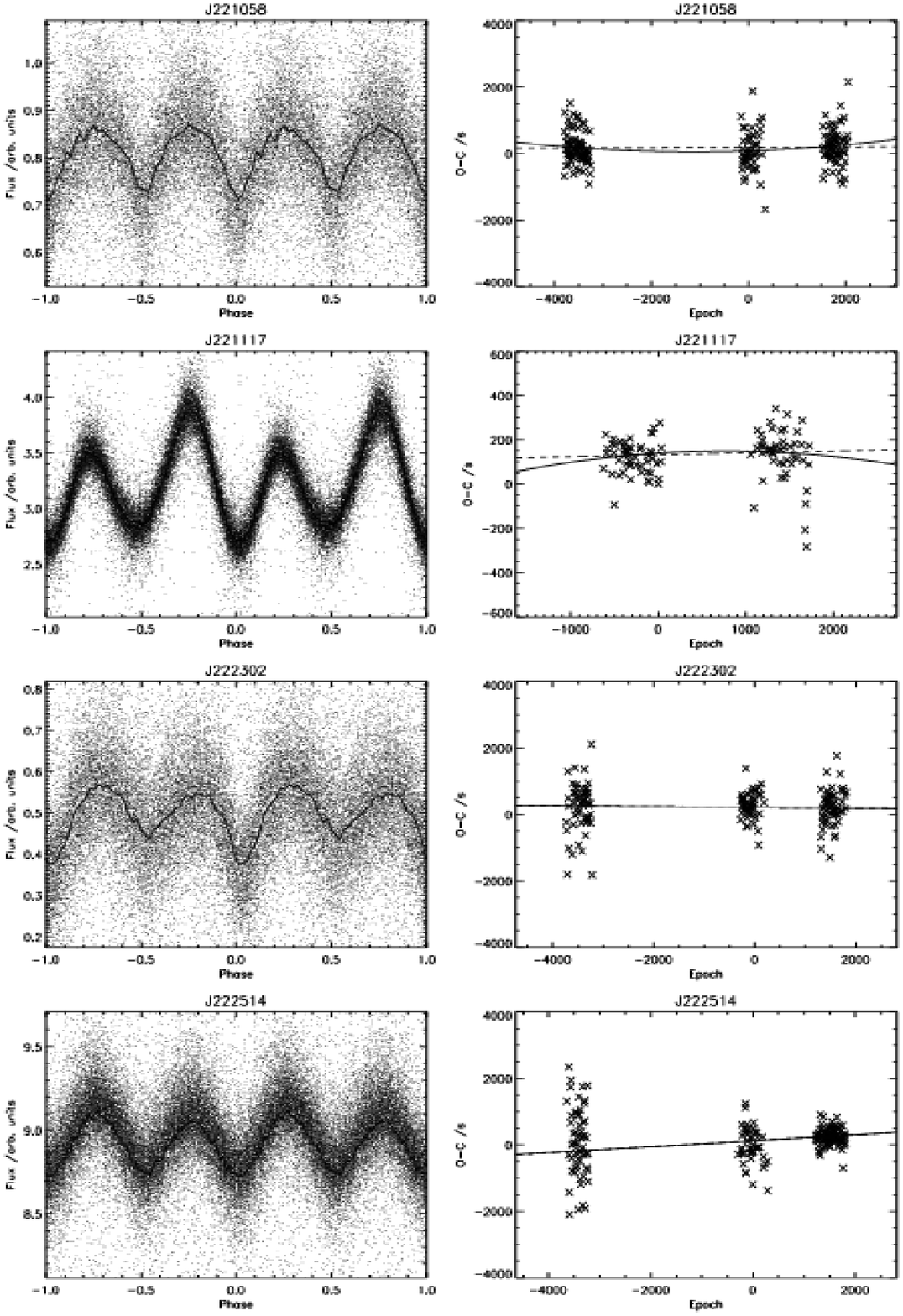}
\end{center}\textbf{Fig.~\ref{appfig1}} continued. \clearpage

\begin{center} \includegraphics[width=16cm]{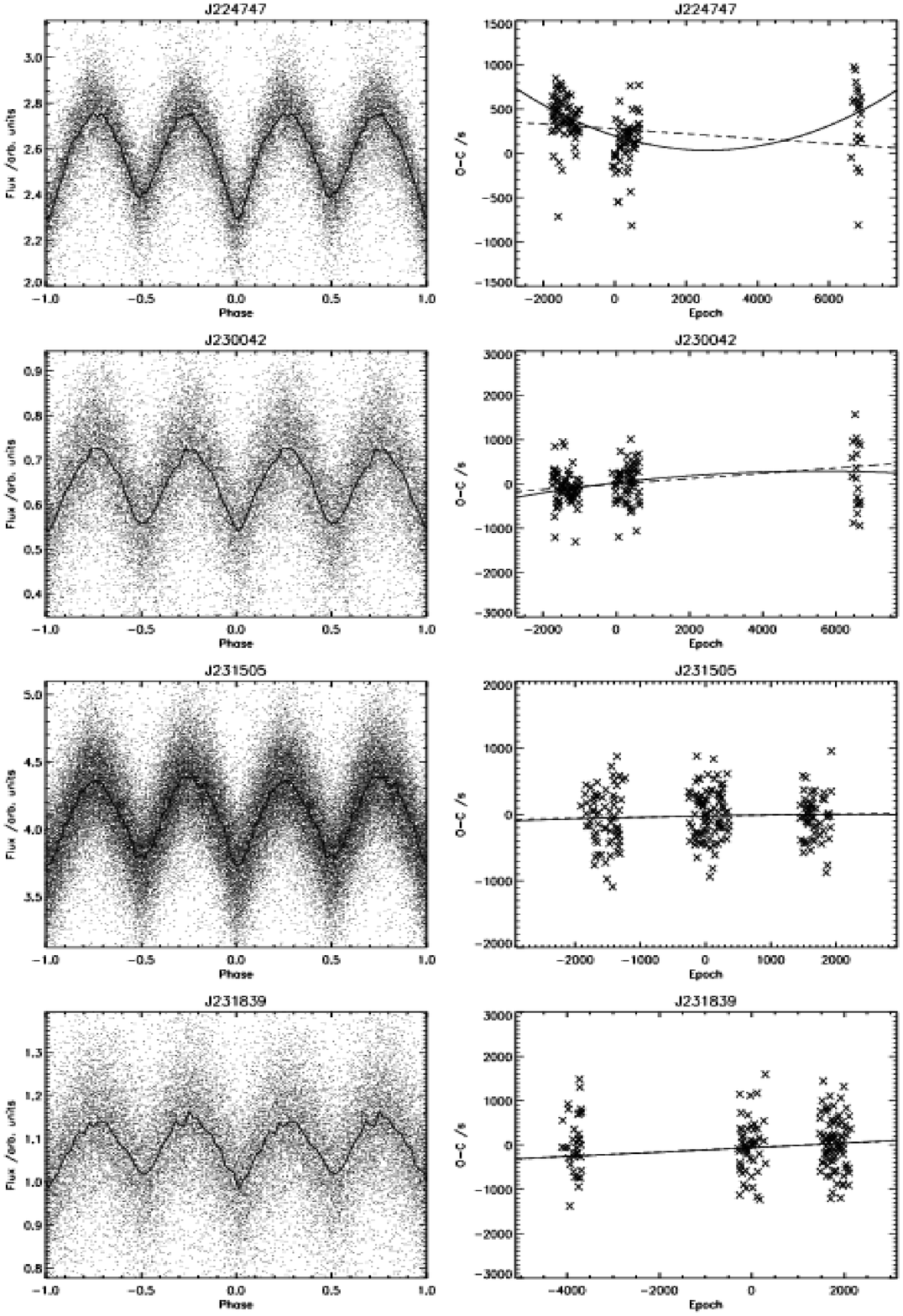}
\end{center}\textbf{Fig.~\ref{appfig1}} continued. \clearpage

\begin{center} \includegraphics[width=16cm]{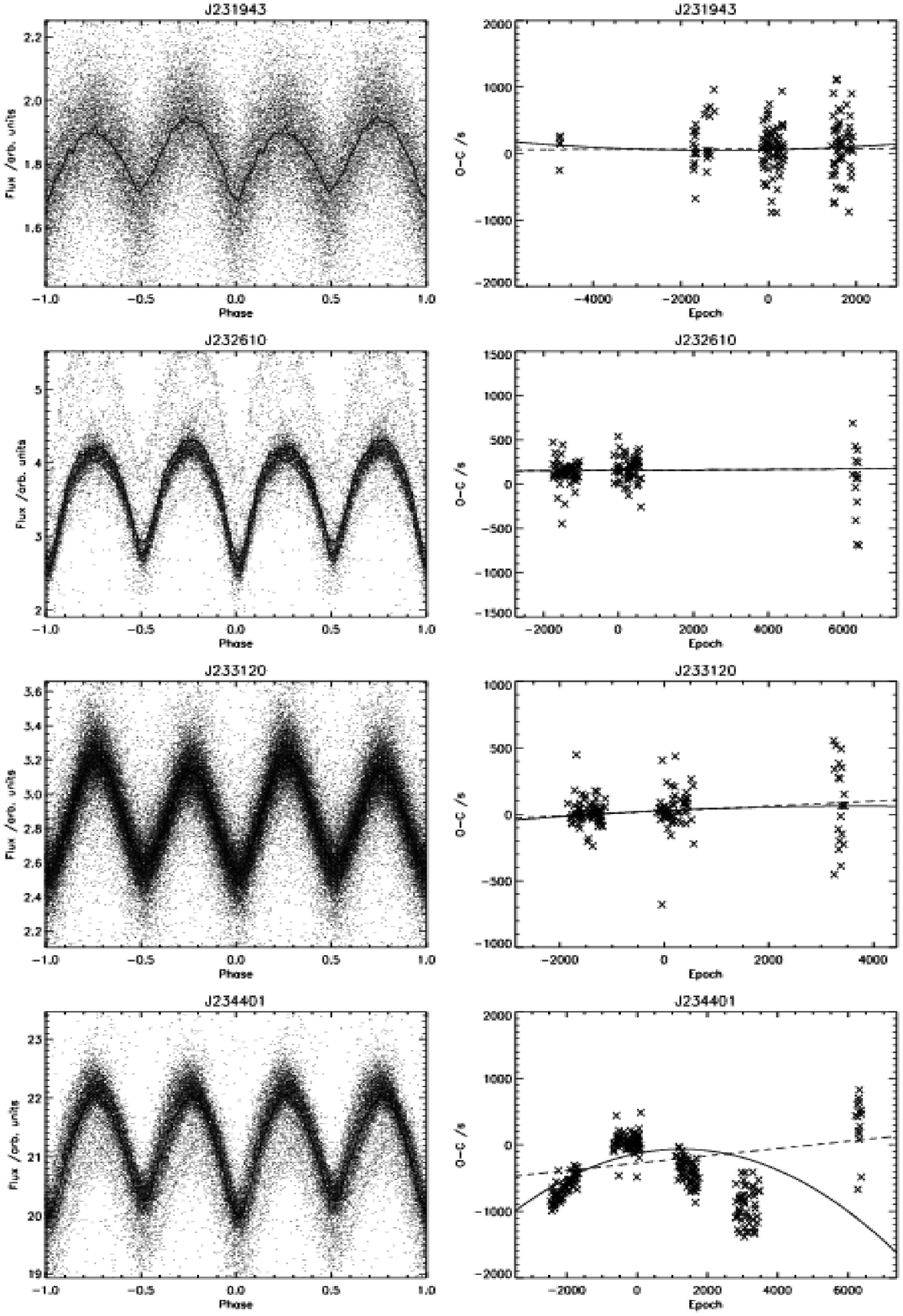}
 \end{center}\textbf{Fig.~\ref{appfig1}} continued. \clearpage

\begin{center} \includegraphics[width=16cm]{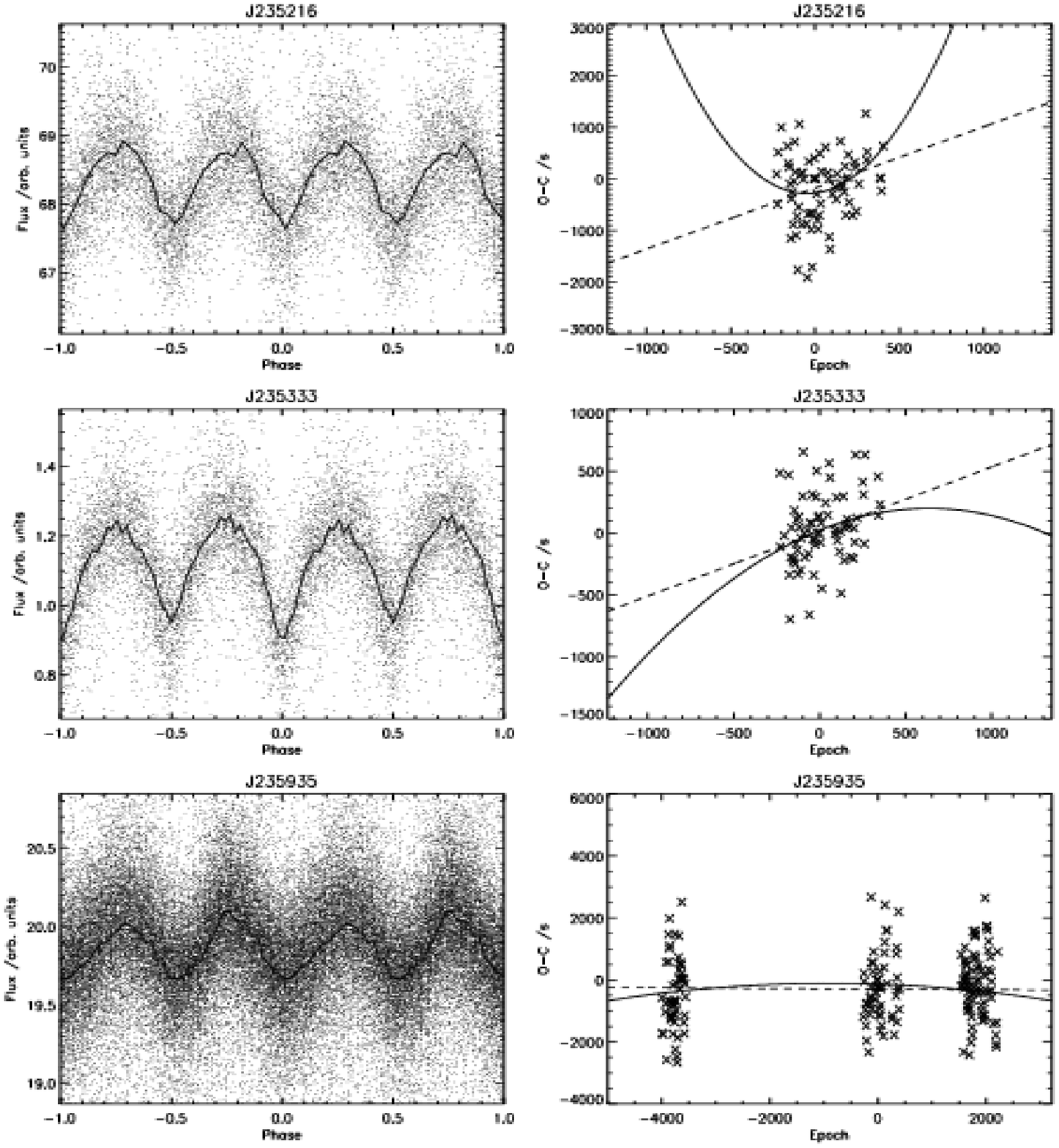}
\end{center}\textbf{Fig.~\ref{appfig1}} continued. \clearpage

\end{document}